\documentclass[twocolumn]{aastex631}
\usepackage{hhline, amsmath}
\usepackage{graphicx}
\usepackage{xspace, times}
\usepackage{hyperref}
\usepackage{color, comment}

\newcommand{\um}{\,$\mu$m\xspace}

\newcommand{\chandra}{\emph{Chandra}\xspace}

\shorttitle{AGN in SPOGs}
\shortauthors{Lanz et al.}

\begin{document}

\title{Are Active Galactic Nuclei in  Post-Starburst Galaxies \\Driving the Change or Along for the Ride?}

\author[0000-0002-3249-8224]{Lauranne Lanz}
\affiliation{Department of Physics, The College of New Jersey, 2000 Pennington Road, Ewing, NJ 08628, USA}

\author[0000-0002-7850-7093]{Sofia Stepanoff}
\affiliation{Department of Physics, The College of New Jersey, 2000 Pennington Road, Ewing, NJ 08628, USA}

\author[0000-0003-1468-9526]{Ryan C. Hickox}
\affiliation{Department of Physics and Astronomy, Dartmouth College, 6127 Wilder Laboratory, Hanover, NH 03755, USA }

\author[0000-0002-4261-2326]{Katherine Alatalo}
\affiliation{Space Telescope Science Institute, 3700 San Martin Dr, Baltimore, MD 21218, USA}
\affiliation{Johns Hopkins University, The William H. Miller III Department of Physics \& Astronomy, Baltimore, MD 21218, USA}
\author[0000-0002-4235-7337]{K. Decker French}
\affiliation{Department of Astronomy, University of Illinois Urbana-Champaign, Urbana, IL 61801, USA}
\affiliation{Center for Astrophysical Surveys, National Center for Supercomputing Applications, \\1205 W. Clark St., Urbana, IL 61801, USA}
\author[0000-0001-7883-8434]{Kate Rowlands}
\affiliation{AURA for ESA, Space Telescope Science Institute, 3700 San Martin Dr, Baltimore, MD 21218, USA}
\affiliation{Johns Hopkins University, Department of Physics and Astronomy, Baltimore, MD 21218, USA}
\author[0000-0003-1991-370X]{Kristina Nyland}
\affiliation{U.S. Naval Research Laboratory, 4555 Overlook Ave. SW, Washington, DC 20375, USA}

\author[0000-0002-7607-8766]{Phil Appleton}
\affiliation{IPAC, Mail Code 314-6, Caltech, 1200 E. California Blvd., Pasadena, CA 91125, USA}
\author[0000-0002-3032-1783]{Mark Lacy}
\affiliation{National Radio Astronomy Observatory, 520 Edgemont Road, Charlottesville, VA 22903, USA}
\author[0000-0001-7421-2944]{Anne Medling}
\affiliation{Ritter Astrophysical Research Center, University of Toledo, Toledo, OH 43606, USA}
\affiliation{ARC Centre of Excellence for All Sky Astrophysics in 3 Dimensions (ASTRO 3D)}
\author[0000-0003-2083-5569]{John S. Mulchaey}
\affiliation{The Observatories of the Carnegie Institution for Science, Pasadena, CA 91101, USA;}
\author[0000-0001-6245-5121]{Elizaveta Sazonova}
\affiliation{Johns Hopkins University, Department of Physics and Astronomy, Baltimore, MD 21218, USA}
\author[0000-0002-0745-9792]{Claudia Megan Urry}
\affiliation{Yale Center for Astronomy \& Astrophysics, 46 Hillhouse Avenue, New Haven, CT 06511, USA}
\affiliation{Department of Physics, Yale University, P.O. Box 208120, New Haven, CT 06520-8120, USA}

\begin{abstract}

We present an analysis of 10\,ks snapshot \chandra observations of 12 shocked post-starburst galaxies, which provide a window into the unresolved question of active galactic nuclei (AGN) activity in post-starburst galaxies and its role in the transition of galaxies from actively star forming to quiescence. While 7/12 galaxies have statistically significant detections (with 2 more marginal detections), the brightest only obtained 10 photons. Given the wide variety of hardness ratios in this sample, we chose to pursue a forward modeling approach to constrain the intrinsic luminosity and obscuration of these galaxies rather than stacking. We constrain intrinsic luminosity of obscured power-laws based on the total number of counts and spectral shape, itself mostly set by the obscuration, with hardness ratios consistent with the data. We also tested thermal models. While all the galaxies have power-law models consistent with their observations, a third of the galaxies are better fit as an obscured power-law and another third are better fit as thermal emission. If these post-starburst galaxies, early in their transition, contain AGN, then these are mostly confined to a lower obscuration ($n_H \leq10^{23}$\,cm$^{-2}$) and lower luminosity ($L_{2-10\,\rm keV}\leq10^{42}$\,erg\,s$^{-1}$). Two galaxies, however, are clearly best fit as significantly obscured AGN. At least half of this sample show evidence of at least low luminosity AGN activity, though none could radiatively drive out the remaining molecular gas reservoirs. Therefore, these AGN are more likely along for the ride, having been fed gas by the same processes driving the transition.

\end{abstract}

 \section{Introduction}
 
 Galaxies in the local Universe show correlated bimodal distributions in color, star formation activity, and morphology, with a majority falling into either a blue actively star-forming spiral category or a red quiescent lenticular or elliptical category \citep[e.g.,][]{strateva01, baldry04}. The relative rarity of galaxies with intermediate colors, the ``green valley'' galaxies, suggests a relatively rapid transition \citep{bell03}. However, green valley galaxies, if selected based just on color, contain a mix of two populations: a smaller rapidly quenched post-starburst population and a much larger slower quenching population whose optical colors shift with the build-up of an older stellar population and the exhaustion of their molecular gas \citep[e.g.,][]{noeske07, schawinski14}. 
 
Selecting these rapid quenchers during their transition therefore requires spectral criteria, yielding populations called `E+A' or `K+A' galaxies \citep[e.g.,][]{Zabludoff96, Quintero04, Goto07}. This methodology selects for recent star formation activity on the basis of H$\delta$ absorption associated with A stars, while excluding galaxies showing ionized emission lines associated with active star formation. This exclusion inevitably also removes galaxies with other energetic processes, which could be involved in the quenching process such as emission from active galactic nuclei (AGNs). Several new selection criteria have since emerged that are not as strict in excluding ionized emission \citep[e.g., ][]{wild07, yesuf14, alatalo16a_spog1}. The galaxies used in this study were selected via the \citet{alatalo16a_spog1} methodology and belong to a population referred to as shocked post-starburst galaxies (SPOGs). The SPOGs selection retains galaxies with ionized emission consistent with shocks, AGNs, and low-ionization nuclear emission regions  \citep[LINERs;][]{kauffmann03, kewley06} but excludes those that fall within the star-forming region of the three line diagnostic diagrams \citep{baldwin81, veilleux87}. The SPOGs selection typically selects galaxies earlier in their transition than the classical E+A criteria, as discussed further below. As such, they likely contain more readily observable clues as to the process driving the transition to quiescence. 

The SPOGs sample was initially presented by \citet{alatalo16a_spog1}, where the sample selection methodology was discussed in detail. This sample was also compared to the sample of classical post-starburst galaxies collected by \citet{Goto07}, particularly in the color-magnitude space, demonstrating that SPOGs are generally bluer than the Goto post-starburst galaxies. As such, the inclusion of a larger range of gas emission line ratios as compared to emission line cuts tends to select galaxies that have on-average younger stellar populations and are  therefore earlier in their transition  than classically-selected post-starburst galaxies. Similarly, \citet{ardila18} found that SPOGs have bluer FUV-NUV colors than other post-starburst samples. 

In \citet{french18}, the star formation histories of SPOGs were determined together with those of post-starburst galaxies which were selected using cuts on H$\delta$ absorption and absence of H$\alpha$ emission. In this study, UV and optical photometry together with optical spectroscopy were fit with stellar population models to obtain star formation histories. Both sets of post-starburst galaxies showed similar histories and stellar mass distributions consistent with other post-starburst samples (e.g., \citealt{melnick13}), but SPOGs generally had younger ages consistent with being recent progenitors of the comparison post-starburst samples. This result is in agreement with the blue optical and UV colors found by \citet{alatalo16a_spog1} and \citet{ardila18}.

A major open question in galaxy evolution and in particular in the transition from star forming to quiescence is the role of AGN in this process.  Mergers have long been posited as a pathway for this transition \citep[e.g.,][]{toomre72, hopkins06} and are effective at creating both periods of AGN activity \citep[e.g.,][]{spring05b, hopkins10} and post-starburst galaxies in simulations \citep{zheng2020, lotz2020}. AGN feedback has also been shown to be effective as a means of quenching star formation in simulations \citep[e.g.,][]{springel05a, somerville08}, including in field post-starburst galaxies specifically \citep{lotz2020}.  AGNs are also known to drive multiphase outflows that have the potential to strip their galaxies of star-forming gas \citep[e.g.,][]{sturm11, cicone14} and inject shocks and turbulence into their host's interstellar media \citep[ISM; e.g.,][]{alatalo15, lanz16, smercina18} thereby reducing the star formation activity. As such, AGN feedback provides a viable means of explaining the quenching in post-starburst galaxies or the lack of the resumption of star formation in the remaining molecular gas \citep{ rowlands15, french15, alatalo16b_spog2}.

However, the observational support for both of these ingredients, mergers and AGN feedback, in post-starburst galaxies is mixed. Depending on the study, merger fractions in these galaxies can range from 15\% to 70\%   \citep{Zabludoff96, goto05, yang08, pawlik16, Sazonova21}. The lower end of this range is likely an underestimate due to fading signatures of mergers over time. However, alternative processes may also play a factor, particularly for galaxies in denser environments. 

 Similarly, observational evidence for the presence and activity of AGN in post-starburst galaxies remains unclear with a variety of results. \citet{brown09} found that roughly half of 24 K+A galaxies in the NOAO Deep Wide-Field Survey showed optical signatures consistent with AGN or LINERs and a third of the optically brighter galaxies had X-ray luminosities of $\sim10^{42}$\,erg\,s$^{-1}$, indicating a relatively high presence of AGN activity, if at a lower activity level. The sample selection was also explicitly biased against emission-line galaxies, thereby excluding AGNs and potentially underestimating the AGN fraction. The presence of AGNs in post-starburst galaxies is further supported by the stacking study undertaken by \citet{georgakakis08}, which found evidence of a significant population of obscured AGN in these galaxies. \citet{yesuf14} also found that 36\% of quenching post-starbursts showed optical signatures of AGNs, albeit delayed relative to the starburst peak, indicating that AGNs were likely not the cause of the quenching in most of these systems. \citet{Depropris14} showed that their sample of 10 K+A galaxies in a range of merger states lacked evidence of powerful AGNs using both optical and X-ray indicators, but a large fraction of their sample had weak ($L_X\leq 10^{40}$\,erg\,s$^{-1}$) X-ray emission. 

\begin{deluxetable*}{lclccccccccr}
\tabletypesize{\scriptsize}
\tablecaption{Sample\label{Sample}}
\tablewidth{\textwidth}
\centering
\tablehead{
\colhead{IAU} & \colhead{SPOG} & \colhead{CXO} & \colhead{R.A.} &\colhead{Dec.} &  \colhead{Redshift} & \colhead{Dist.} &\colhead{\bf  F$_{1.4}$} & \colhead{MW} & \colhead{ObsID} &\colhead{Obs.} & \colhead{Exposure}\\
\colhead{Name} & \colhead{Number} & \colhead{Name} & \colhead{[hms]} & \colhead{[dms]} & \colhead{} & \colhead{(Mpc)}& \colhead{(mJy)}&\colhead{nH} & \colhead{}  & \colhead{Date} & \colhead{Time} \\
\colhead{(1)} & \colhead{(2)} & \colhead{(3)} & \colhead{(4)} & \colhead{(5)} & \colhead{(6)}  & \colhead{(7)} & \colhead{(8)} & \colhead{(9)}   & \colhead{(10)} & \colhead{(11)}  & \colhead{(12)}  
}
\startdata
J001145--005431  & ~~~4 & 38  & 00:11:45.2 & --00:54:30.6  & 0.0479  & 215  & 2.08 & 3.48 & 19480 & 2017-08-09 & 9.99		\\
J085357+031034 &157   & 10     & 08:53:56.8 & +03:10:33.6  & 0.1292 & 613  & 1.25 & 3.16 & 19486 & 2017-01-11 & 9.91		\\
J091407+375310 &186   & 1562 & 09:14:07.2  &+37:53:10.0 & 0.0719 & 328  & 2.60 & 1.50 & 19484 & 2017-01-11 & 9.93		\\
J093820+181953 & 224  & 1070  & 09:38:19.9 & +18:19:52.7 & 0.0886 & 409 & 4.52 & 2.57 & 19485 & 2017-03-08 & 9.93 		\\
J095750--001239 & 253  & 2178  & 09:57:49.5 & --00:12:39.2 & 0.0330 & 146 & 0.86 & 2.77 & 19479 & 2017-03-30 & 9.91		\\
J102653+434008 & 305  & 1558  & 10:26:53.4 & +43:40:08.4 & 0.1053 &  492 & 1.30 & 1.20 & 19481 & 2018-03-03 & 9.92		\\
J113655+245325 & 462 & 1030   & 11:36:55.2 & +24:53:25.4 &0.0327   & 145 & 2.55 & 1.93 & 19478 & 2017-02-02 & 10.80 		\\
J113939+463132 & 470 & 1557   & 11:39:39.3 & +46:31:32.2& 0.1735  & 847 & 4.95 & 1.97 & 19489/20829 & 2017-10-27 & 4.88/4.89\\
J115341+093026 & 498 & 5260   & 11:53:41.3 & +09:30:25.6 & 0.1389   & 663 & 1.56 & 1.73 & 19483 & 2017-02-10 & 9.93	 \\
J131448+210626 & 662 & 1081  &  13:14:47.6 & +21:06:26.3 & 0.0458  & 205 & 1.97 & 2.75 & 19487 & 2016-11-17 & 9.80 	\\
J132648+192246 & 689 & 1147 & 13:26:48.1 & +19:22:45.7 & 0.1741   & 850 & 2.09 & 1.61 & 19488 & 2017-03-11 & 9.91 		\\
J155525+295551 & 955 & 2      &  15:55:24.9 & +29:55:50.9& 0.0699   & 319 & 2.96 & 2.83 & 19482 & 2016-12-14 & 9.82 		\\
\enddata
\tablecomments{(1): IAU name. (2) ID number from \citet[Table 1]{alatalo16a_spog1} which differs from the earlier numbering used at the time of the \chandra proposal (3), included here for ease of reference relative to the archive. (4) and (5) are J2000 coordinates. (6) SDSS spectroscopic redshift. (7) Distance in Mpc determined from the redshift assuming the WMAP 9 cosmology \citep{hinshaw13}.  (8) Integrated 1.4 GHz from FIRST observations from \citet[Table 1]{alatalo16b_spog2}  (9) Foreground Milky Way column density ($nH$) in units of $10^{20}$\,cm$^{-2}$ \citep{nhtool} (10) \chandra observation ID. (11) Observation date. (12) Exposure time in ks. }
\end{deluxetable*}

\begin{figure*}[ht]
\centerline{\includegraphics[trim=0.cm .0cm 0.cm 0cm, clip, width=\linewidth]{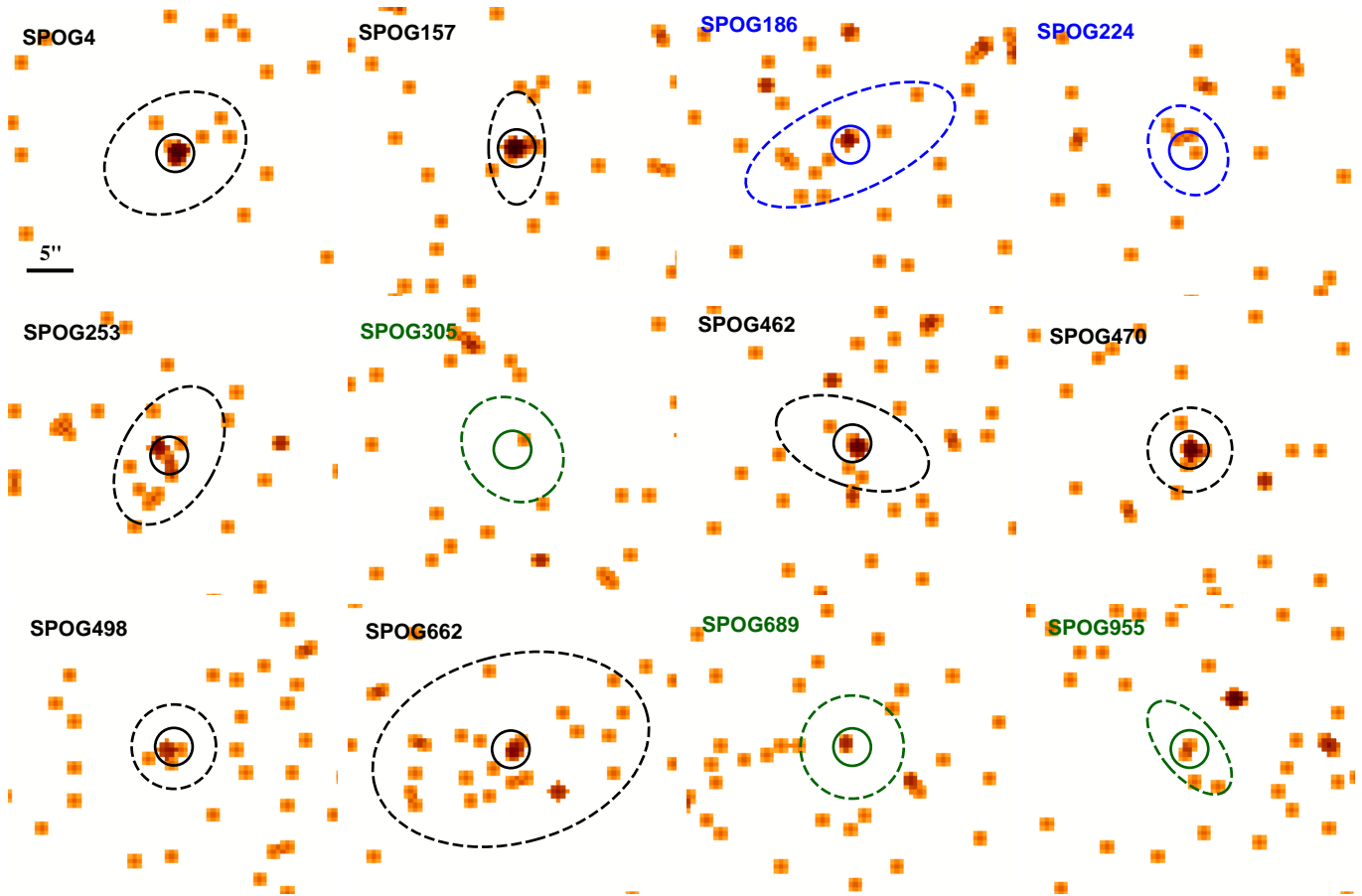}}
\caption{Images of the 0.5--8 keV emission of the 12 SPOGs with the 2$''$ aperture overlaid. The dashed ellipse is the approximate extent of the SDSS {\em r} emission, denoting the size of the galaxy. The color of the label and regions indicates whether the detection is significant (black), marginal (blue), or not-significant (green). Slight Gaussian smoothing is applied to help highlight the presence of the source. }
\label{images}
\end{figure*}

Surveys at radio and infrared wavelengths found less support for the presence of AGN. Cross-identification with the Faint Images of the Radio Sky at Twenty centimeters \citep[FIRST;][]{becker95} survey has been done for two samples of E+A galaxies \citep{shin11, Nielsen12}, but while some enhanced radio emission was detected, it could not be unambiguously identified as AGN activity. More recently, two surveys of the infrared emission of E+A galaxies \citep{alatalo17_psb, Meusinger17} found enhanced mid-IR emission potentially suggestive of AGN but not meeting the WISE color criterion indicative of luminous AGNs \citep[e.g.,][]{stern12}.  While \citet{melnick13} found that the mid-IR emission could be explain by thermally pulsing asymptotic giant branch (TP-AGB) stars or obscured AGNs, \citet{alatalo17_psb} showed that WISE [12]-[22] colors of the Goto post-starburst galaxies are more consistent with AGNs than TP-AGB stars.  \citet{alatalo16b_spog2} also examined the WISE emission in a sample of SPOGs with measurements or limits on molecular masses. In comparison to the correlation between molecular mass and infrared luminosity found in star forming samples, effectively a reframing of the Kennicutt-Schmidt \citep[e.g.,][]{Kennicutt98} relation of constant star formation efficiency, SPOGs showed an excess of IR emission for their molecular content, as could be expected if additional heating is due to an AGN. However, the relatively small excess compared to radio galaxy AGNs suggests a conclusion similar to those of the E+A studies, namely that the IR properties indicate either that AGNs are not especially common or that they are generally low luminosity or radiatively inefficient. 

To make progress determining whether AGNs play a significant role in the transition process, we need an unambiguous measure of AGN activity. Therefore, we obtained \chandra observations of a set of 12 SPOGs selected from the original sample of 1067 \citep[see][for a discussion of the SPOG selection criteria]{alatalo16a_spog1} to have a FIRST counterpart, detected 22$\mu$m emission, and a CO detection with either CARMA or IRAM \citep{alatalo16b_spog2}. These galaxies were among the most likely SPOGs to contain an AGN and the CO measurements allowed at least an initial estimate of obscuration column. We present the results of this investigation here. 
 
This paper is organized as follows. In Section 2, we describe the observations and their analysis. We explain the forward modeling methodology we used to constrain AGN properties in Section 3. The results of this analysis are discussed in Section 4, and their implication are explained in Section 5. Finally, we summarize the paper in Section 6. We assume the {\em Wilkinson Microwave Anisotropy Probe} (WMAP) 9 cosmology with $H_0 = 69.3$\,km/s/Mpc, $\Omega_\Lambda=0.713$, and $\Omega_M=0.287$ \citep{hinshaw13}.

\section{Observations and Data Analysis}
 
 Table \ref{Sample} summarizes our observations (P.I. L. Lanz). Each galaxy was observed for $\sim$10\,ks between 2016-11-17 and 2018-03-03, mostly in a single observation. Each of the galaxies were centered on the aimpoint of the back-illuminated S3 chip of the \chandra Advanced CCD Imaging System \citep[ACIS;][]{weisskopf00} in VFAINT mode. We reprocessed the observations using CIAO version 4.13 to create new level 2 event files. We measured counts in the total (0.5--8 keV), hard (2--8 keV), and soft (0.5--2 keV) bands in 2$''$ apertures centered on the galaxy as well as a larger background region within the S3 chip.  For SPOG470, we measured the counts on each event file and combined them to get the counts  obtained in the full $\sim$10\,ks. Table \ref{counts} lists the results, and Figure \ref{images} shows the aperture placed on each total image.

\subsection{Determination of Statistical Significance \label{prob}}

Since we are clearly in the realm of small number statistics, we use Poisson statistics to determine the probability of getting the observed number of counts given the background level \citep{Gehrels86}. The Poisson probability of getting fewer than or equal to $x$ events given an expected rate of $b$ set by the background estimated for the aperture is
\begin{equation}
P(x,b) = \sum_{k=0}^{x} e^{-b} \frac{b^k}{k!}.
\end{equation}
To determine the probability of the photons in the aperture being background emission, what we need is the probability of getting at least $x$ events, which is therefore given by
\begin{equation}
P_2(x,b) = 1-\sum_{k=0}^{x-1} e^{-b} \frac{b^k}{k!},
\end{equation}
since detections of photons are discrete events. The typical estimated background level in the source aperture is $\sim0.3$ counts in the 0.5--8 keV band, of which $\sim0.1$ come from the 0.5--2 keV band and the other $\sim0.2$ come from the 2--8 keV band. For simplicity of interpretation, we then convert these probabilities to Gaussian $\sigma$ values using
\begin{equation}
P_2(x,b) = {\rm erf}\Big(\frac{\sigma}{\sqrt{2}}\Big).
\end{equation}
Probabilities consistent with $\sigma\geq3$ are considered statistically significant. Those with probabilities less than 1\% of being background fluctuations, corresponding to $\sigma\gtrsim2.6$, are considered marginal detections. The $\sigma$ values for the full 0.5--8 keV band are reported in Table \ref{counts}.

 \begin{deluxetable*}{lccccccc}
\tabletypesize{\scriptsize}
\tablecaption{Counts and Hardness Ratios\label{counts}}
\tablewidth{\textwidth}
\centering
\tablehead{
\colhead{IAU} & \colhead{SPOG} & \multicolumn{3}{c}{(Net) Counts} &  \colhead{Significance} & \colhead{Hardness} \\
\cline{3-5}\\
\colhead{Name} & \colhead{Number} & \colhead{0.5-8 keV} & \colhead{0.5-2 keV} & \colhead{2-8 keV} & \colhead{0.5-8 keV}  & \colhead{Ratio}  \\\\
\colhead{(1)} & \colhead{(2)} & \colhead{(3)} & \colhead{(4)} & \colhead{(5)} & \colhead{(6)}  & \colhead{(7)}  
}
\startdata
J001145--005431 & ~~~4 & 8 (7.7) & 3 (2.9) & 5 (4.8) & 6.2 & 0.24 ([-0.07, 0.58])\\ 
J085357+031034 &157  & 10 (9.7)~\, & 1 (0.9) & 9 (8.8) & 7.3 & 0.79 ([0.69, 1.00])\\ 
J091407+375310 &186  & 3 (2.7) & 1 (0.9) & 2 (1.8) & 2.9$^\star$ & 0.30 ([-0.01, 1.00])\\ 
J093820+181953 & 224 & 3 (2.7) & 2 (1.9) & 1 (0.8) & 2.9$^\star$ & -0.40 ([-1.00, -0.12])\\ 
J095750--001239 & 253 & 6 (5.7) & 5 (4.9) & 1 (0.8) & 4.8 & -0.72 ([-1.00, -0.56])\\ 
J102653+434008 & 305 & 1 (0.7) & 1 (0.9) & 0 (0)~~~ & 1.1$^\dagger$ & -0.64 ([-1.00, -0.27])\\ 
J113655+245325 & 462 & 5 (4.7) & 4 (3.9) & 1 (0.8) & 4.3 & -0.65 ([-1.00, -0.47])\\
J113939+463132 & 470 & 7 (6.7) & 6 (5.9) & 1 (0.8) & 5.7 & -0.74 ([-1.00, -0.60])\\ 
J115341+093026 & 498 & 4 (3.7) & 2 (1.9) & 2 (1.8) & 3.6 & -0.05 ([-0.53, 0.48])\\ 
J131448+210626 & 662 & 4 (3.7) & 2 (1.9) & 2 (1.8) & 3.7 & -0.05 ([-0.50, 0.49])\\ 
J132648+192246 & 689 & 2 (1.7) & 1 (0.9) & 1 (0.8) & 2.1$^\dagger$ & -0.05 ([-1.00, 0.29])\\ 
J155525+295551 & 955 & 2 (1.8) & 1 (0.9) & 1 (0.8) & 2.3$^\dagger$ & -0.09 ([-1.00, 0.24]) 
\enddata
\tablecomments{(3)--(5) Total counts and background-subtracted net counts in parentheses in the total, soft, and hard bands. (6) Statistical significance of the total net count detection. See Section \ref{prob} for an explanation of their derivation.  Marginal detections are marked with a $^\star$. Non-significant detections are marked with a $^\dagger$. (7) Median hardness ratio with 1$\sigma$ range given in parentheses from BEHR \citep{park06}. }
\end{deluxetable*}

\subsection{Hardness Ratio Determination}

Due to the low photons counts, we chose to use the Bayesian Estimation of Hardness Ratios \citep[BEHR;][]{park06} to determine the likely range of hardness ratios consistent with our data. Hardness ratios provide a sense of the spectral shape, particularly in the absence of sufficient counts for spectral fitting. We use 0.5--2 keV as our soft band ($S$) and 2--8 keV as the hard band ($H$) with the hardness ratio defined as
\begin{equation}
HR = \frac{H-S}{H+S}
\end{equation}
BEHR takes the total source counts in each band as well as the total background counts in each band and a ratio of the background aperture area to the source aperture area. A Bayesian approach coupled with Monte Carlo sampling is used to determine the posterior distribution for HR. In Table \ref{counts}, we report the median of that distribution in addition to the 1$\sigma$ range.

\subsection{Why Not Stack?}

A common approach when faced with a sample of similar objects with few counts is to stack the spectra and fit it to get an average sense of the X-ray properties of these objects. Although these are all post-starburst galaxies, we decided that this route would provide limited useful information for two main reasons. First, even including the marginal and non-detections, we only had a total of 55 counts, of which approximately 1.2 are from background (with only the significant detections, we only have 44 counts). This stacked spectrum would therefore provide very poor constraints on any obscuration as well as the slope of the intrinsic power-law. 

Second and more importantly, while the X-ray emission is indeed measured from the center of the galaxies, it is unclear whether this is always due to power-law AGN emission in these sources. This concern emanates from the much deeper 150\,ks observation \chandra observation undertaken on the local post-starburst galaxy NGC\,1266\footnote{NGC1266 is a specifically interesting comparison for SPOGs because it has optical emission line characteristics that would have been selected against by classical post-starburst selection criteria but not the SPOGs selection criteria \citep{alatalo16a_spog1}.} (\citealt{alatalo15};  Lanz et al. in prep.). The majority of NGC\,1266's X-ray emission, while concentrated within a few arcseconds of the AGN, is clearly extended, soft, best modeled as a thermal plasma,  and  coincident with a multiphase outflow. Its AGN appears to be both highly obscured and relatively low in luminosity. As such, NGC\,1266 is a clear example of a post-starburst galaxy whose X-ray emission is not dominated by power-law emission. Additionally, radio galaxies, including local early-type galaxies, have been shown to have significant thermal X-ray emission due to shocks driven by their jets, as also traced by warm molecular hydrogen emission \citep{lanz15}. Since the SPOG selection specifically allows galaxies with emission-line ratios consistent with shocks, some of the selected galaxies may have X-ray emission more consistent with thermal emission than power-law emission.

\begin{figure*}
\centerline{\includegraphics[trim=0.cm .0cm 0.cm 0cm, clip, width=\linewidth]{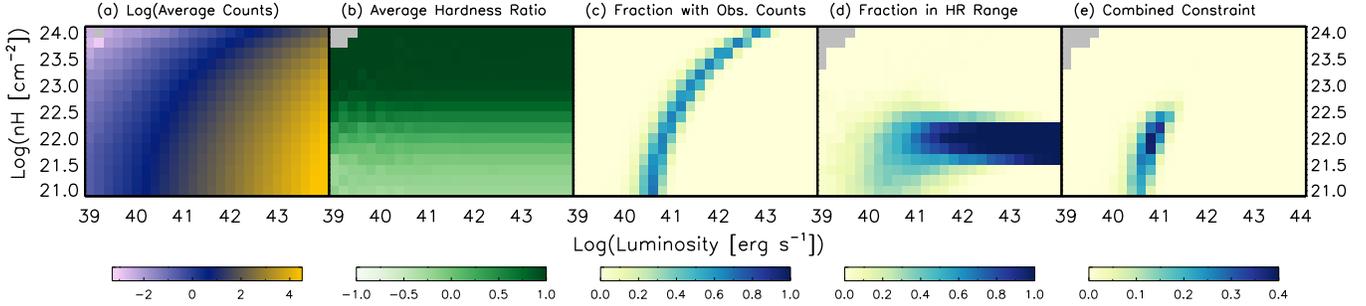}}
\caption{Example of the analysis of a set of power-law models, in this case for SPOG4. The X-axis has the intrinsic 2--10 keV luminosity of the AGN, while the Y-axis has the local obscuration column. (a) The average number of counts in the 0.5--8 keV band for each model, colored with a log scale. Low obscuration, high luminosity models yield the most counts. (b) The average hardness ratio of the spectra for each model. Higher obscuration models have harder spectra. (c) The weighted fraction of spectra for each model with counts consistent with the observed number for SPOG4 (see Section \ref{method} for details). As is to be expected, more luminous models are necessary to get a certain photon flux when the obscuration is higher. (d) The fraction of spectra for each model with hardness ratios within the range given in Table \ref{counts} (column 7). More luminous models yield more counts and so less scatter in their associated hardness ratios. (e) The fraction of spectra per model that meet both the photon flux (panel c) and hardness ratio (panel d) constraints. The gray squares indicates models where less than 5 spectra of the 1000 have non-zero counts and therefore the model does not really have a good measure of its hardness ratio.}
\label{workflowPL}
\end{figure*}

While the hardness ratios of this sample are not tightly constrained, their median values have a large range from the very soft at $-0.72$ (SPOG253) to the very hard $0.79$ (SPOG157). Further, they approximately divide into thirds as hard, soft, or intermediate hardness sources. This suggests either a mix of power-law-dominated and thermal-dominated systems and/or a variety of obscuration levels from minimal to high. Therefore, stacking such a heterogenous group would not provide reliable constraints.

 \section{Forward Modeling Methodology}\label{method}
 
To obtain constraints on the nuclear properties of these galaxies, we therefore turned to forward modeling. We have two main constraints: the number of counts observed and their hardness ratios. The first effectively constrains the normalization of any model as a brighter object will yield more counts in a given exposure time. The second provides a proxy of spectral shape. To generate mock spectra for each galaxy, we needed {\em Chandra}'s response files (rmf and arf) at the time of the observation as well as the exposure time. We used \texttt{sherpa} \citep{sherpa, sherpaPython} to extract response files from each observation\footnote{Since the two observations of SPOG470 were taken on the same day, we used the response of the first observation, as the difference between the response files would be minimal.}, so that the simulated spectra would be consistent with the specific response at the time of the observation of each galaxy. 

For each galaxy, we tested two model types: an absorbed power-law to represent a classical AGN-dominated system and a thermal plasma   as an alternative to represent systems in which outflows or shocked emission dominates. We limit this analysis to these two simple models in order to avoid over-fitting the limited constraints we have, as discussed in the previous paragraph. With effectively two data points per galaxy, our models can have at most two free parameters. This pair of model types spans the range of potential AGN-driven X-ray emission, and a recent study of nearby galaxies \citep{williams22} showed that obscured power-laws with or without a thermal component worked well for a large range of AGN luminosities including Seyfert galaxies and LINERs.  

For the power-law model, these free parameters are the local obscuration ($n_H$), applied with an \texttt{xsphabs} photoelectric absorption model, and an intrinsic 2--10 keV luminosity, used to determine the normalization of the power-law in combination with the distance to the galaxy. The photon index is fixed to 1.8 \citep{piconcelli05, dadina08}. We tested 416 different models consisting of 16 values of local obscuration between $10^{21}$ cm$^{-2}$ and $10^{24}$ cm$^{-2}$ in steps of $10^{0.2}$ and 26 intrinsic luminosities between $10^{39}$ erg s$^{-1}$ and $10^{44}$ erg s$^{-1}$ also in increments of $10^{0.2}$.
\begin{figure*}
\centerline{\includegraphics[trim=0.cm .0cm 0.cm 0cm, clip, width=\linewidth]{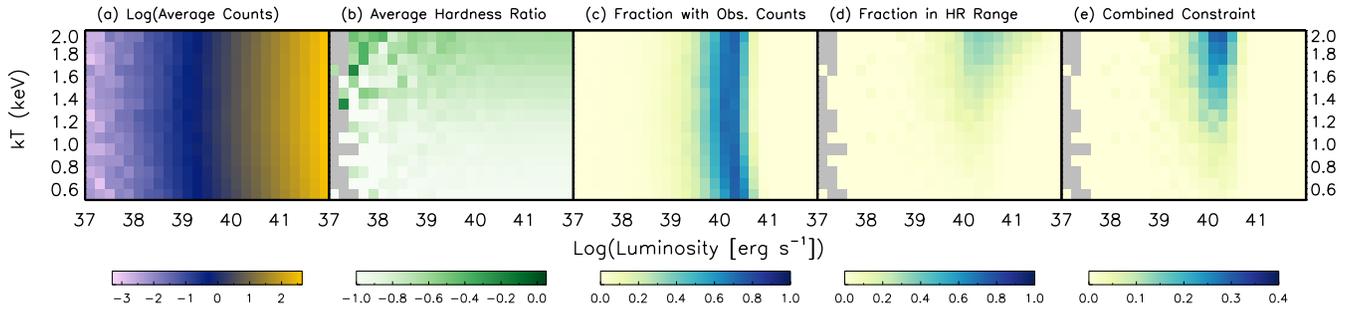}}
\caption{Example of the analysis of a set of thermal plasma models, in this case for SPOG662. The panels are broadly similar to those of Figure \ref{workflowPL}. However, here the X-axis is the 0.5--8 keV luminosity of the plasma and the Y-axis is the temperature of the plasma.  In thermal plasma models, the number of counts (a) does not depend much on the temperature for a given luminosity. However, the temperature sets the hardness ratio (b). Hotter plasma have harder spectra, but all are generally soft. Since the hardness range from \texttt{APEC} models is narrower, it can provide a relatively strong constraint for galaxies with harder spectra but are less effective as a constraint (d) for galaxies with softer emission. The photon flux constraints (c) tend to constrain the 0.5--8 keV luminosity within approximately an order of magnitude.  }
\label{workflowTH}
\end{figure*}

\begin{figure}
\centerline{\includegraphics[trim=0.1cm 2.65cm 12cm 0.65cm, clip, width=0.9\linewidth]{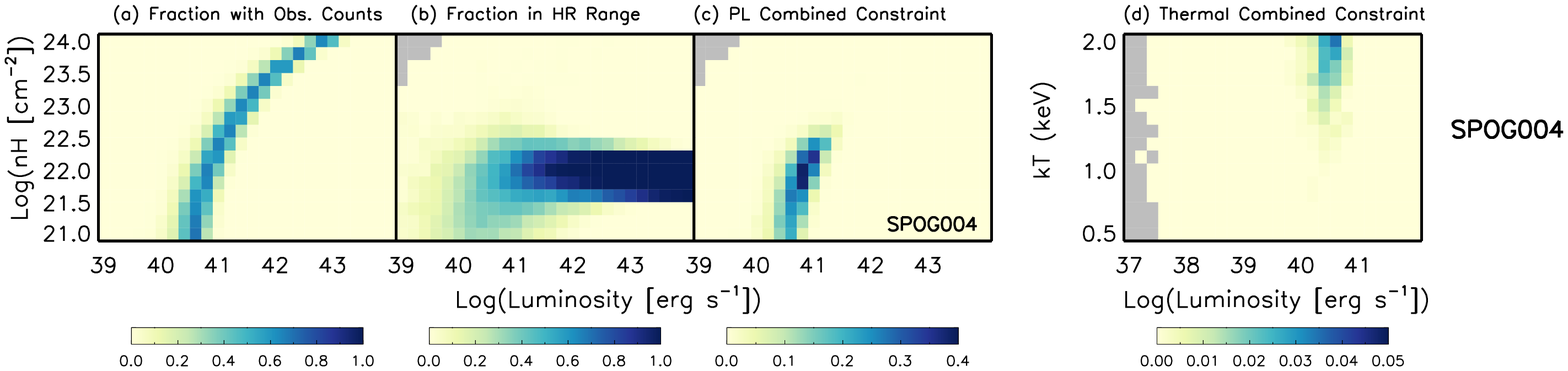}}
\centerline{\includegraphics[trim=0.1cm 2.65cm 12cm 1.12cm, clip, width=0.9\linewidth]{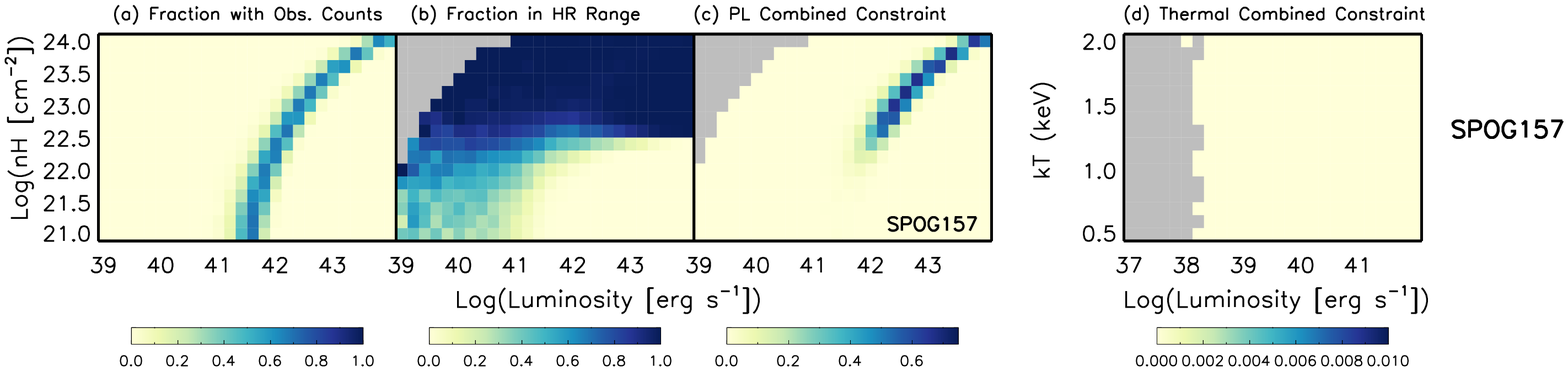}}
\centerline{\includegraphics[trim=0.1cm  2.65cm 12cm 1.12cm, clip, width=0.9\linewidth]{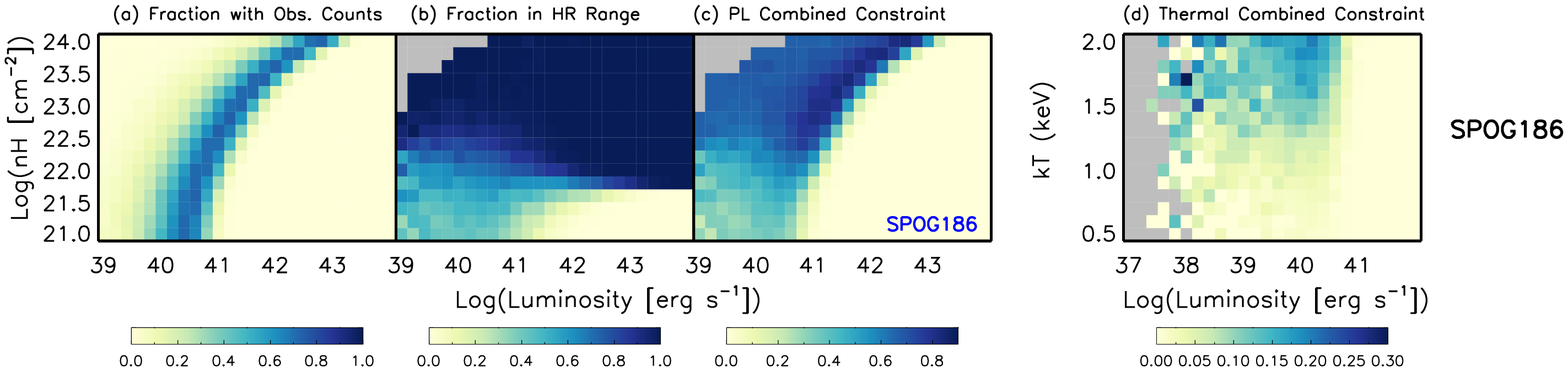}}
\centerline{\includegraphics[trim=0.1cm 2.65cm 12cm 1.12cm, clip, width=0.9\linewidth]{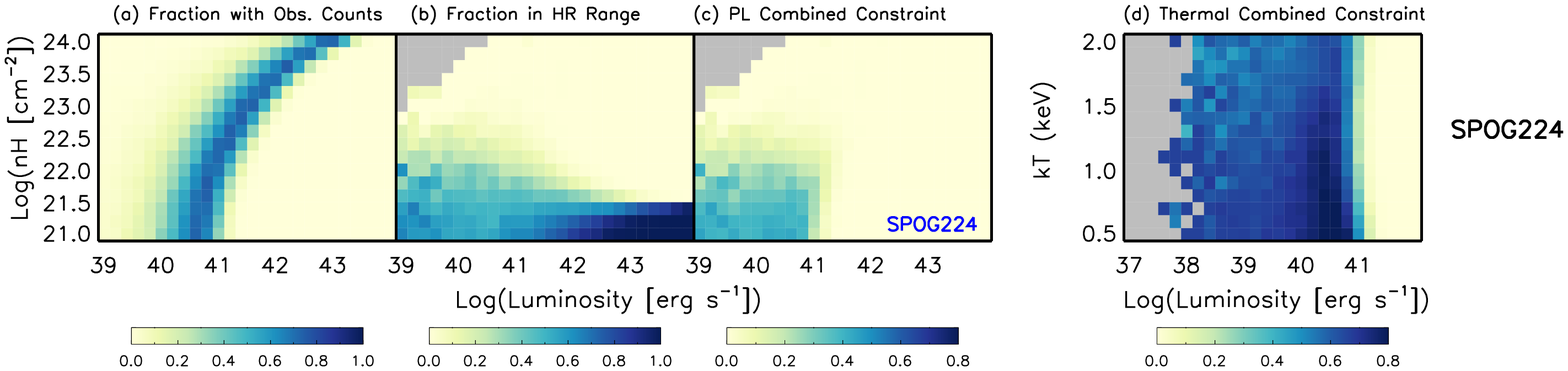}}
\centerline{\includegraphics[trim=0.1cm 2.65cm 12cm 1.12cm, clip, width=0.9\linewidth]{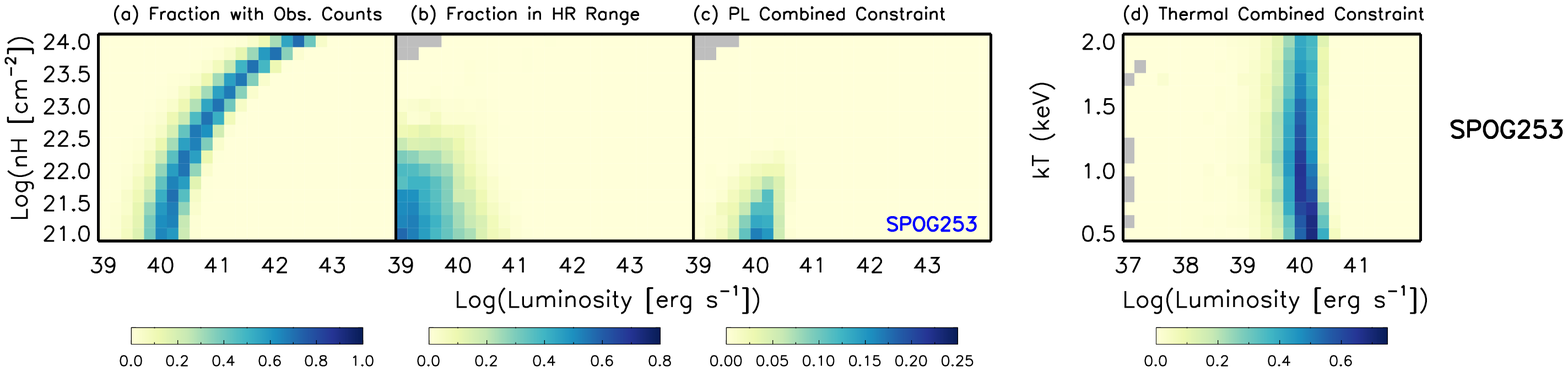}}
\centerline{\includegraphics[trim=0.1cm 2.65cm 12cm 1.12cm, clip, width=0.9\linewidth]{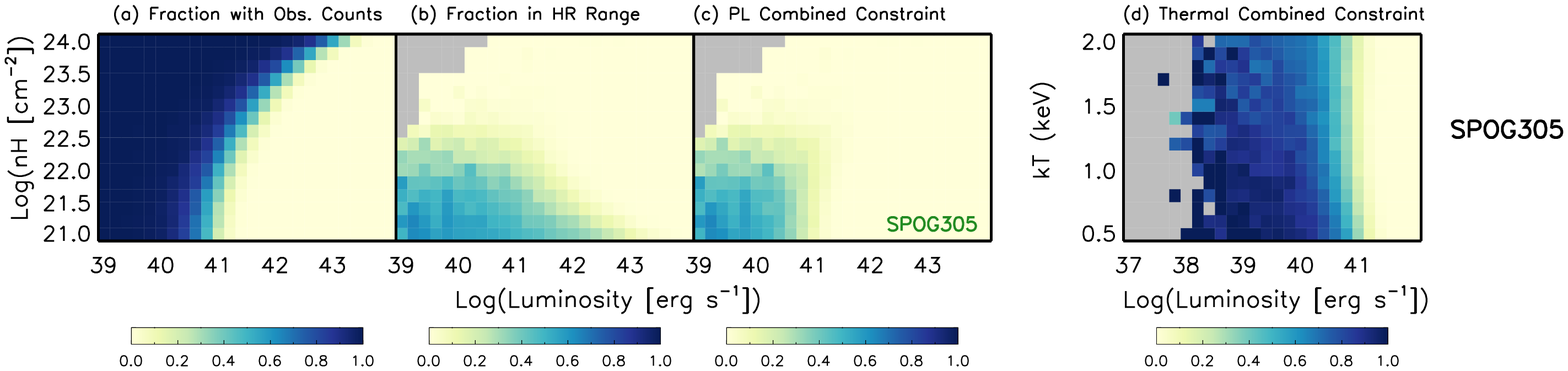}}
\centerline{\includegraphics[trim=0.cm 2.65cm 12cm 1.12cm, clip, width=0.9\linewidth]{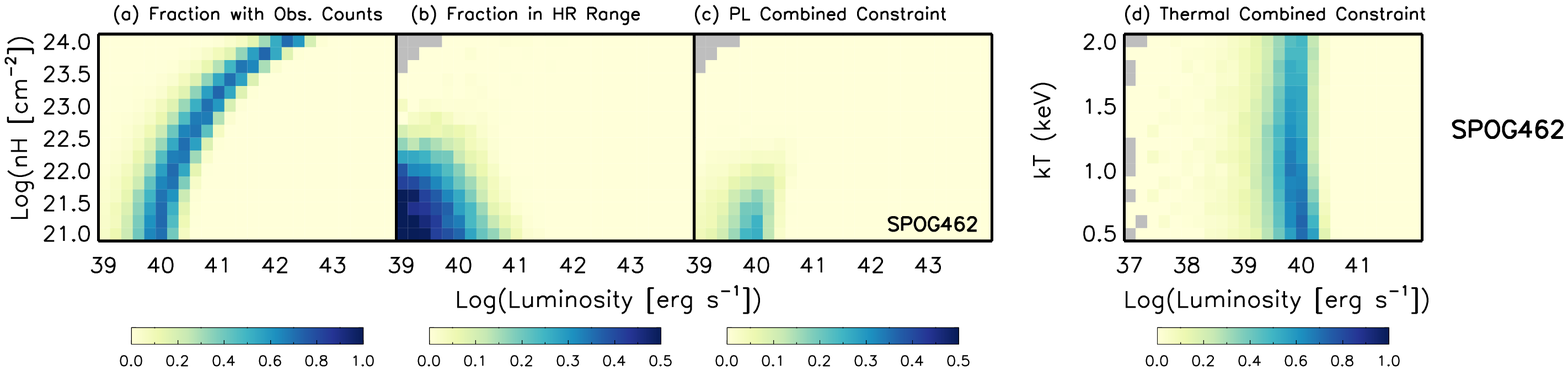}}
\centerline{\includegraphics[trim=0.cm 2.65cm 12cm 1.12cm, clip, width=0.9\linewidth]{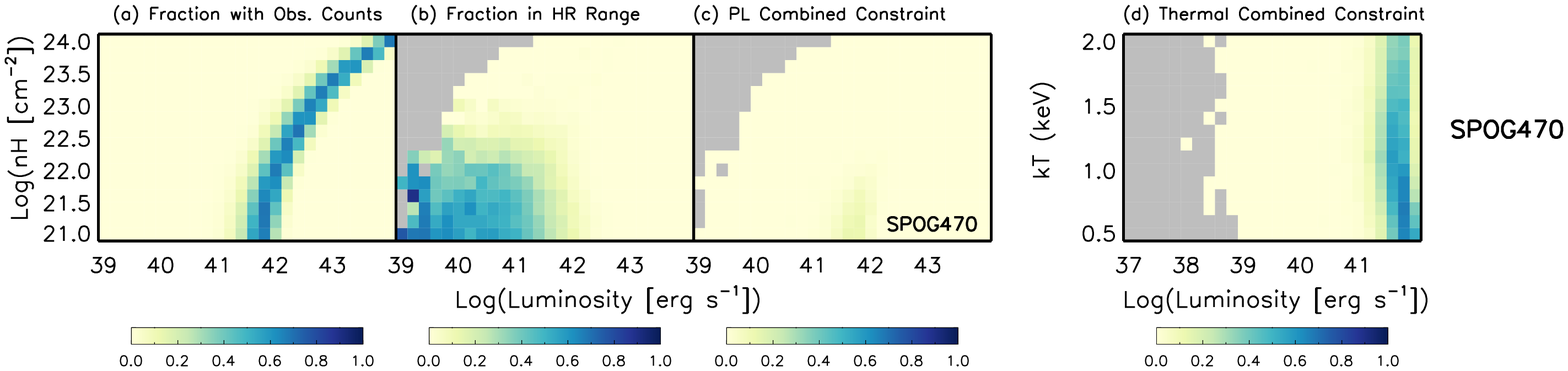}}
\centerline{\includegraphics[trim=0.cm 2.65cm 12cm 1.12cm, clip, width=0.9\linewidth]{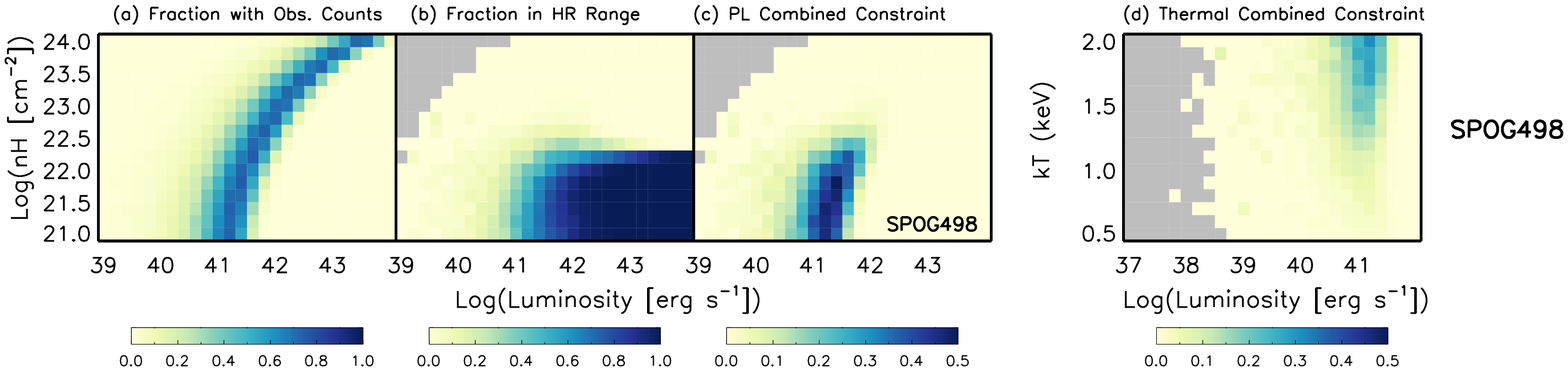}}
\centerline{\includegraphics[trim=0.cm 2.65cm 12cm 1.12cm, clip, width=0.9\linewidth]{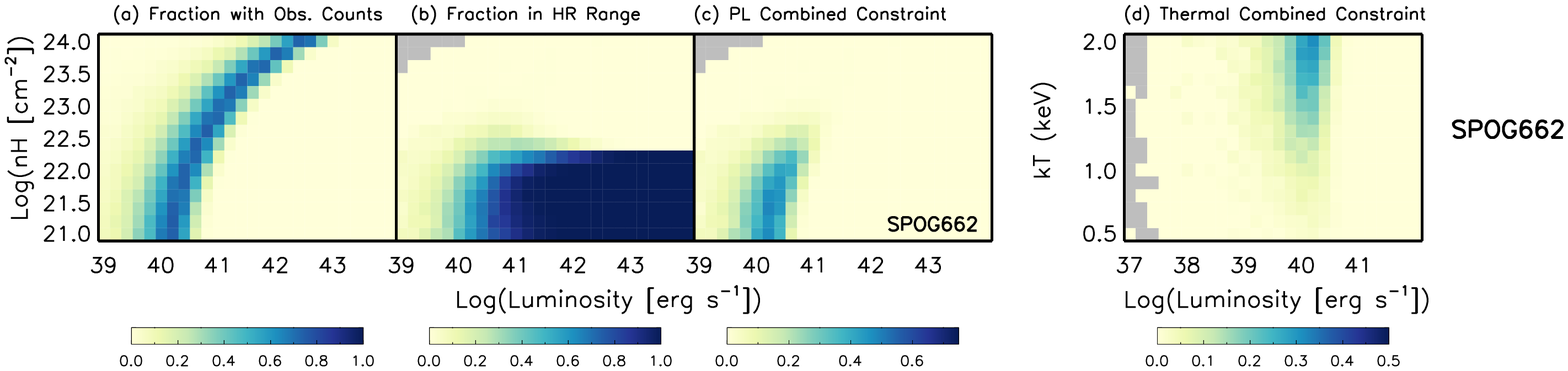}}
\centerline{\includegraphics[trim=0.cm 2.65cm 12cm 1.12cm, clip, width=0.9\linewidth]{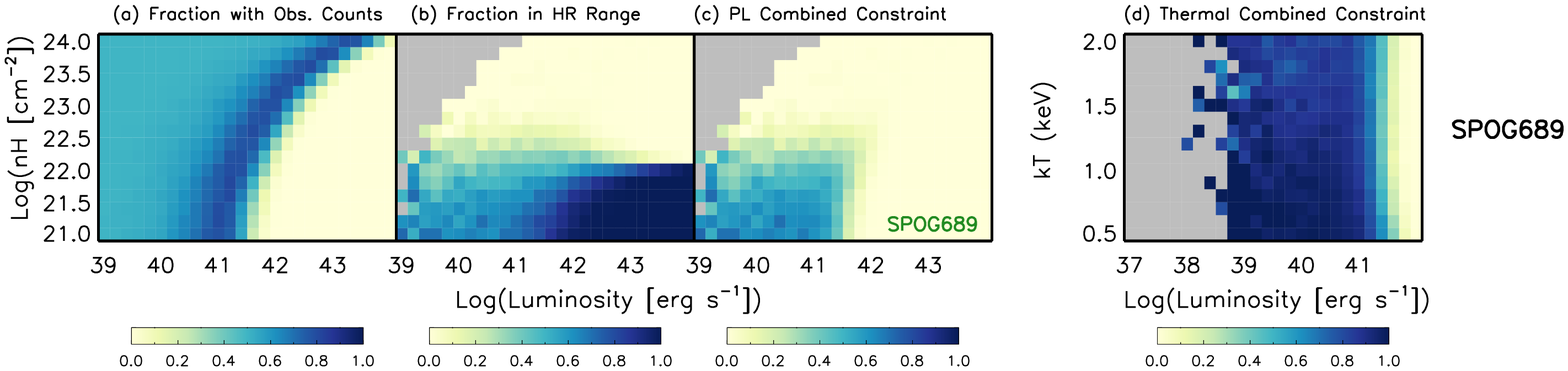}}
\centerline{\includegraphics[trim=0.cm 0.2cm 12cm 1.12cm, clip, width=0.9\linewidth]{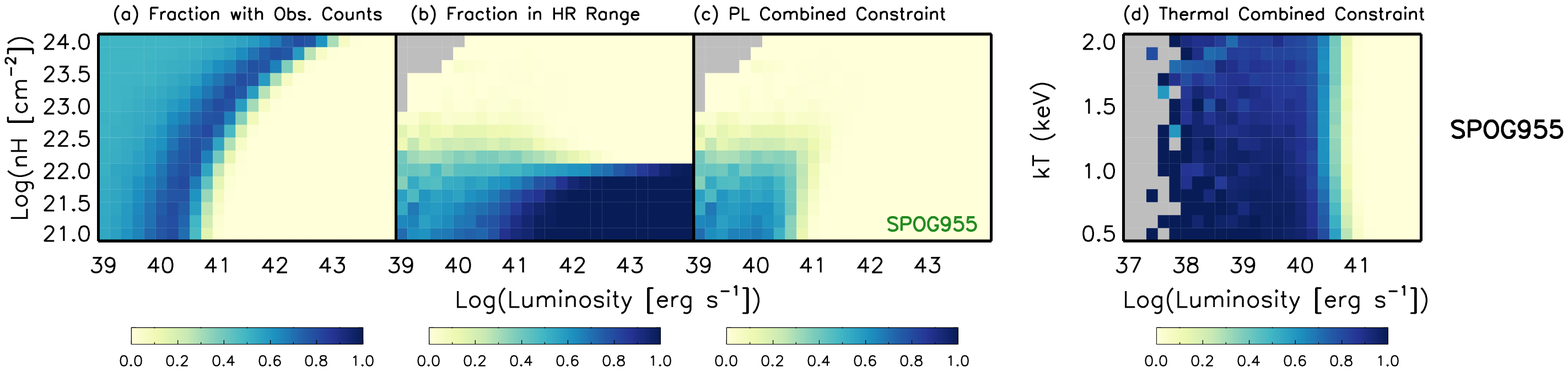}}
\caption{Results of power-law model analysis for each SPOG in numerical order. The left column shows the photon flux constraint, the middle column shows the hardness ratio constraint, and the right column shows the combined constraints. The gray squares are models with fewer than 0.5\% of the spectra with non-zero counts. The color of the galaxy name has the same meaning as in Figure \ref{images}. }
\label{ResultsPL}
\end{figure}

\begin{figure}
\centerline{\includegraphics[trim=0.1cm 2.65cm 12cm 0.65cm, clip, width=0.9\linewidth]{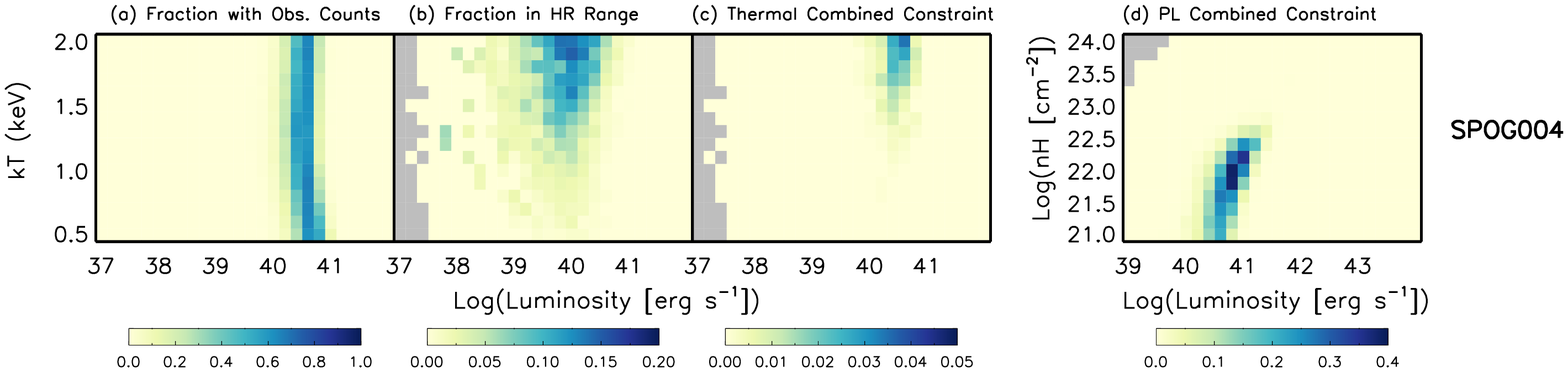}}
\centerline{\includegraphics[trim=0.1cm 2.65cm 12cm 1.12cm, clip, width=0.9\linewidth]{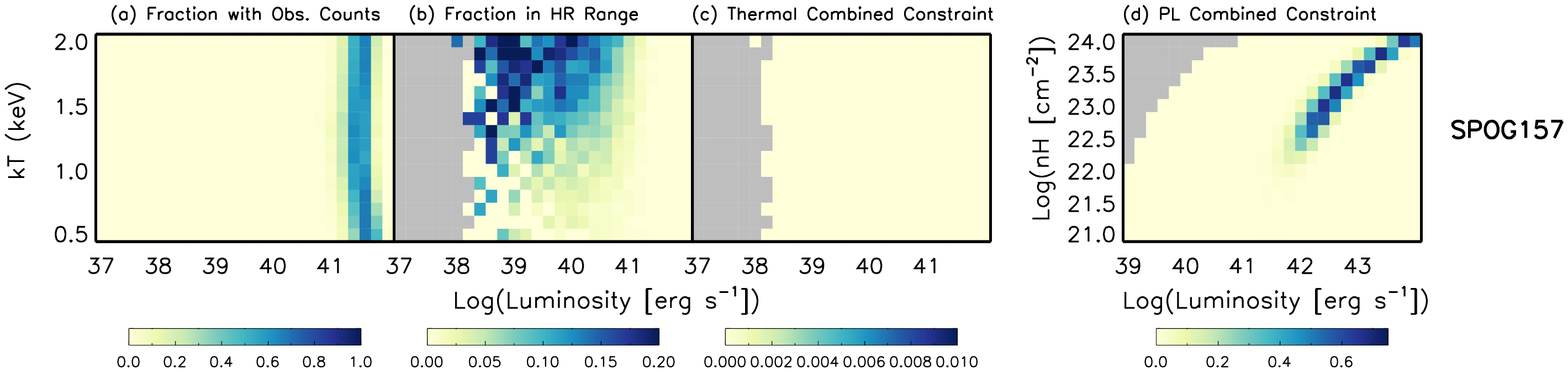}}
\centerline{\includegraphics[trim=0.1cm  2.65cm 12cm 1.12cm, clip, width=0.9\linewidth]{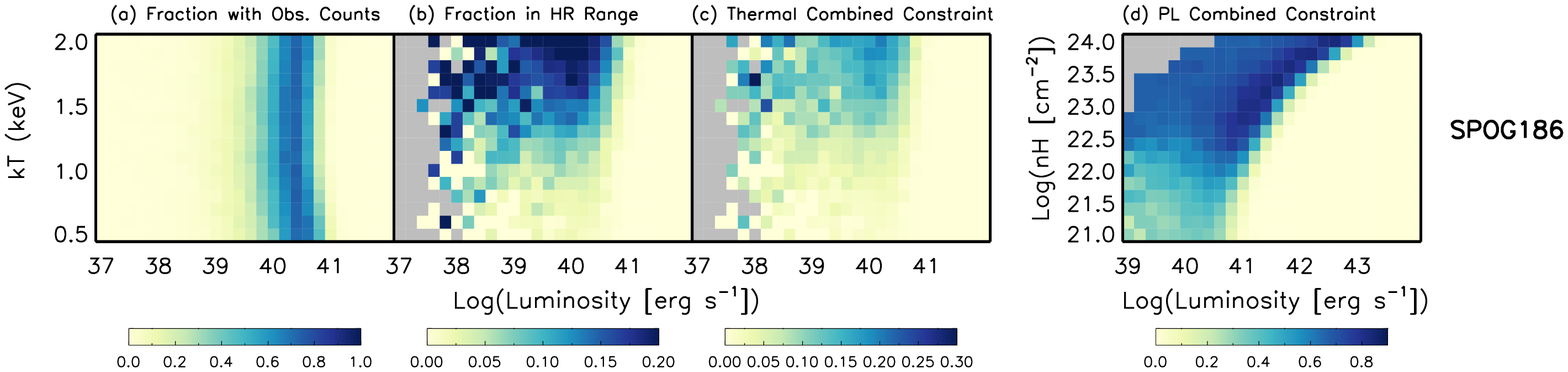}}
\centerline{\includegraphics[trim=0.1cm 2.65cm 12cm 1.12cm, clip, width=0.9\linewidth]{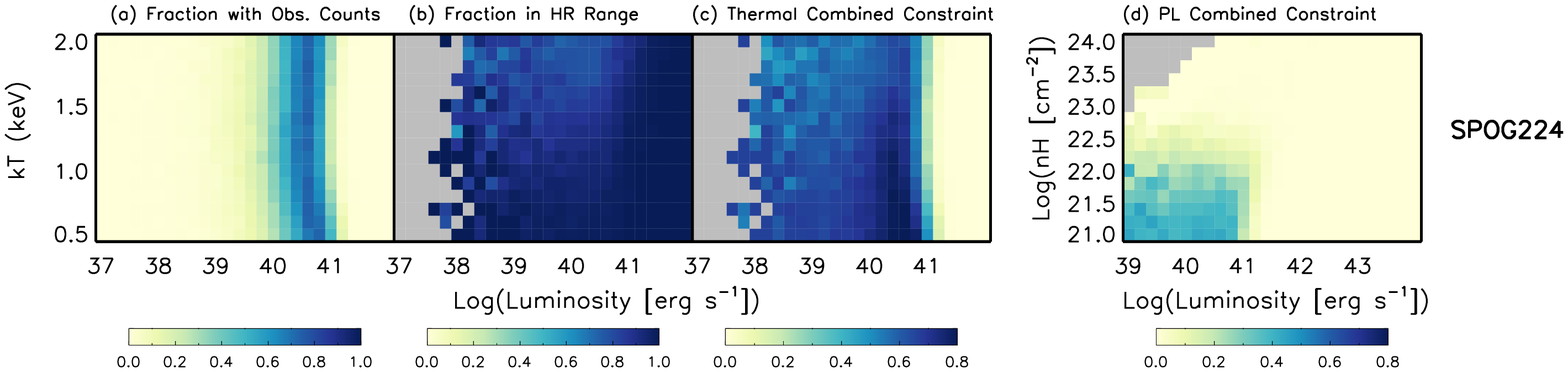}}
\centerline{\includegraphics[trim=0.1cm 2.65cm 12cm 1.12cm, clip, width=0.9\linewidth]{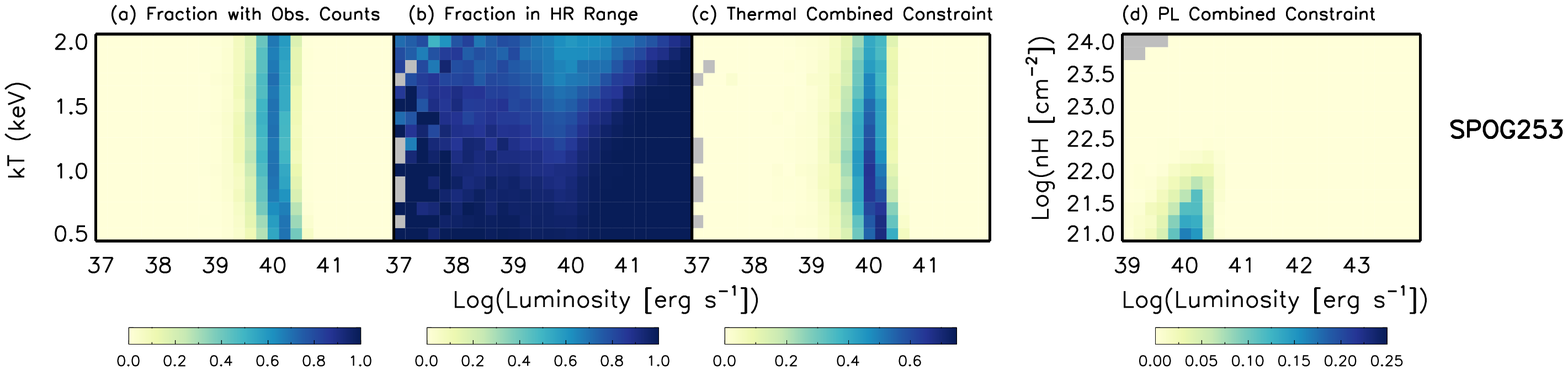}}
\centerline{\includegraphics[trim=0.1cm 2.65cm 12cm 1.12cm, clip, width=0.9\linewidth]{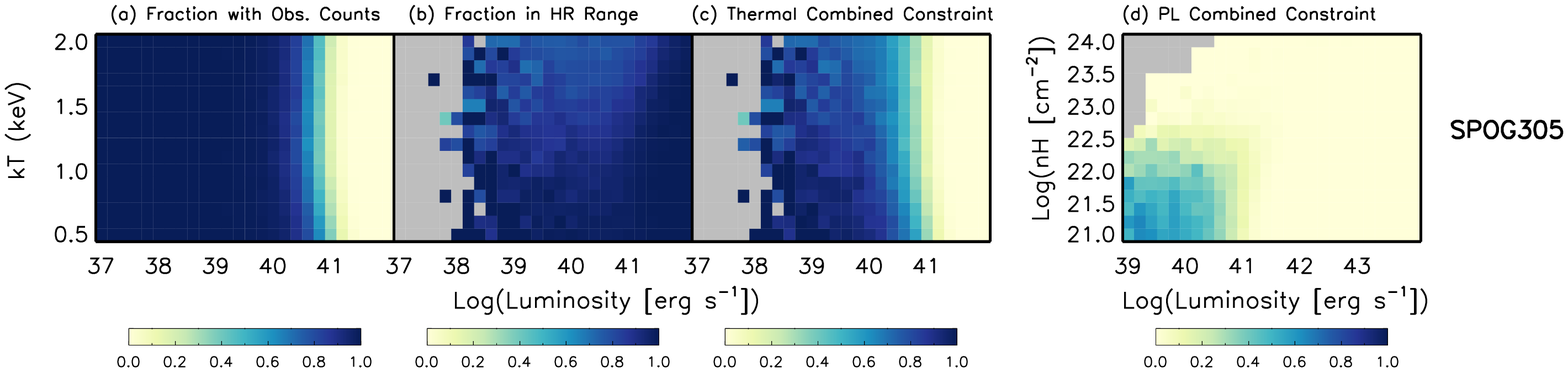}}
\centerline{\includegraphics[trim=0.cm 2.65cm 12cm 1.12cm, clip, width=0.9\linewidth]{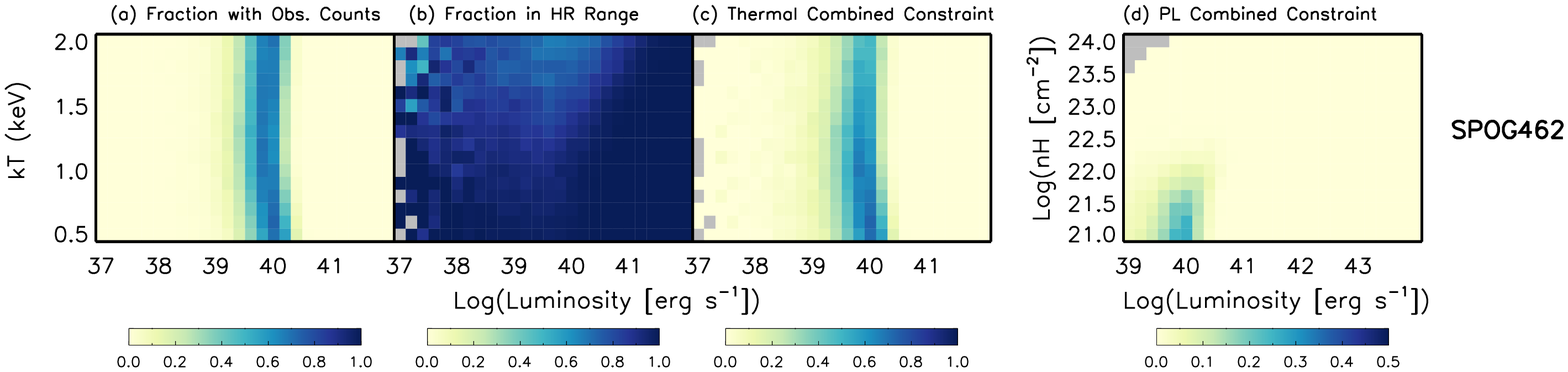}}
\centerline{\includegraphics[trim=0.cm 2.65cm 12cm 1.12cm, clip, width=0.9\linewidth]{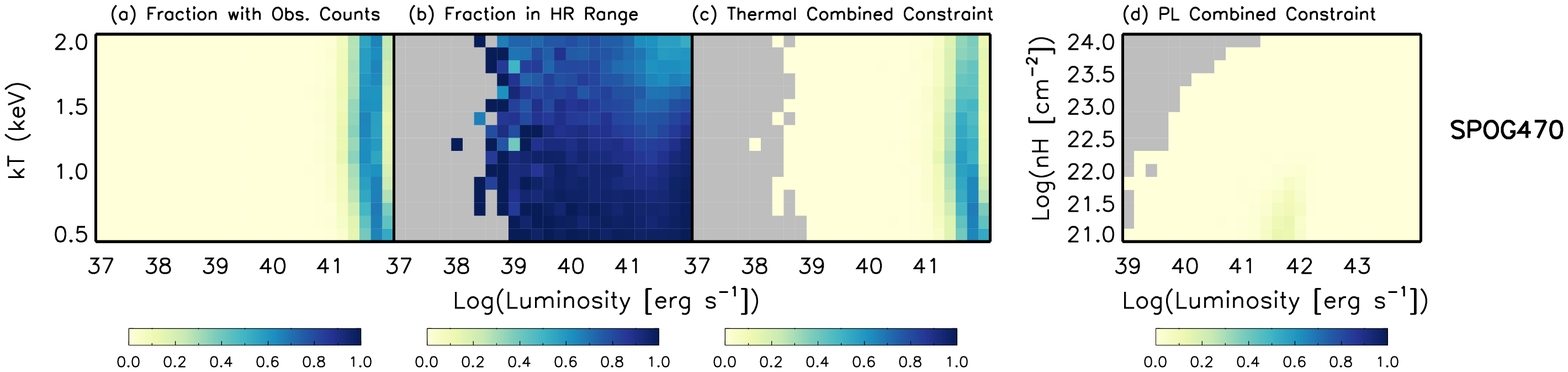}}
\centerline{\includegraphics[trim=0.cm 2.65cm 12cm 1.12cm, clip, width=0.9\linewidth]{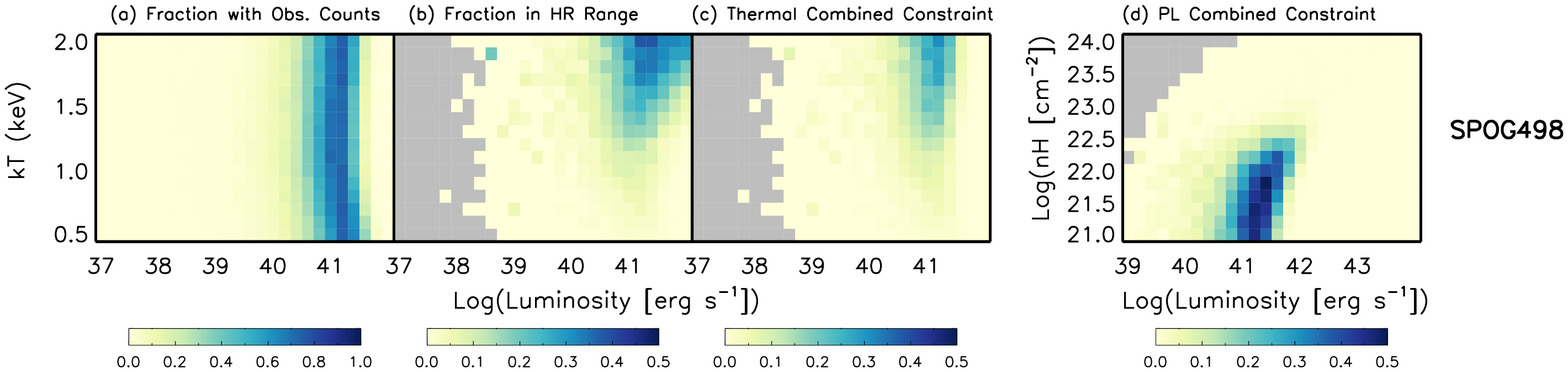}}
\centerline{\includegraphics[trim=0.cm 2.65cm 12cm 1.12cm, clip, width=0.9\linewidth]{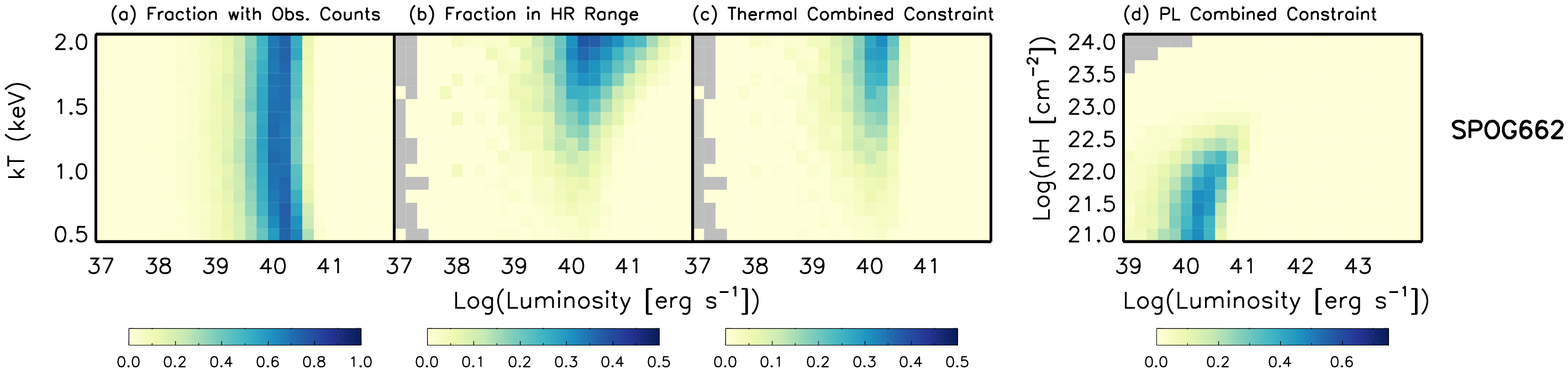}}
\centerline{\includegraphics[trim=0.cm 2.65cm 12cm 1.12cm, clip, width=0.9\linewidth]{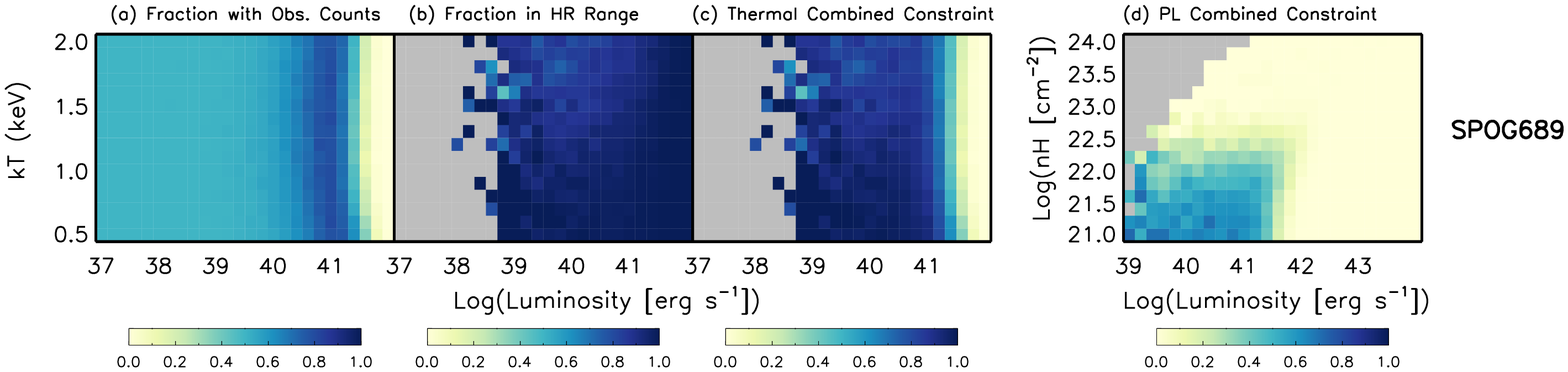}}
\centerline{\includegraphics[trim=0.cm 0.2cm 12cm 1.12cm, clip, width=0.9\linewidth]{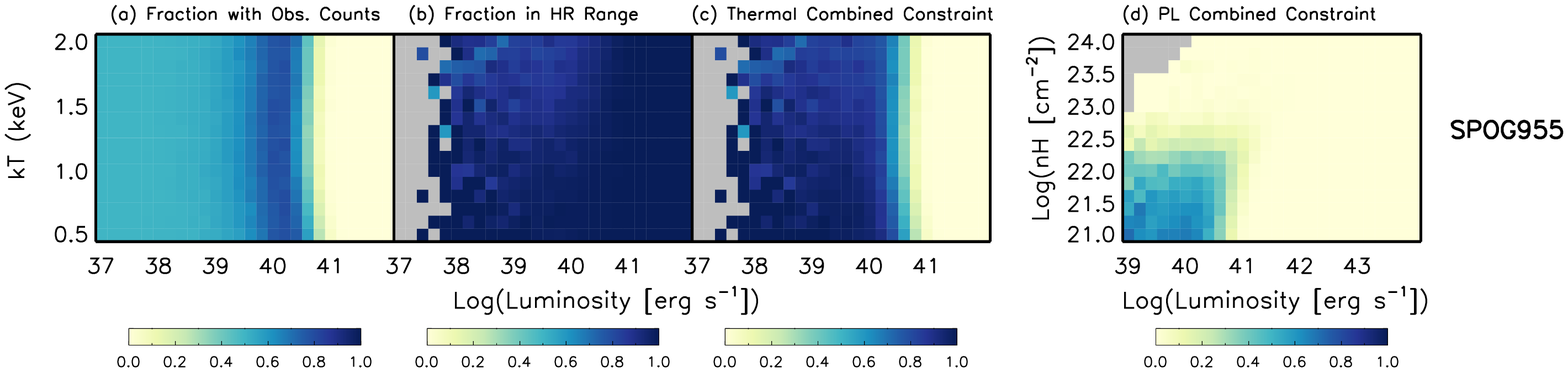}}
\caption{Results of \texttt{APEC} model analysis for each SPOG in numerical order. The left column shows the photon flux constraint, the middle column shows the hardness ratio constraint, and the right column shows the combined constraints. The gray squares are models with fewer than 0.5\% of the spectra with non-zero counts. For more distant galaxies, this corresponds to higher luminosities. }
\label{ResultsTH}
\end{figure}

 The free parameters of the thermal model are the plasma temperature ($kT$) of an \texttt{APEC} \citep{Smith01} model and an intrinsic 0.5--8 keV luminosity, which sets the normalization of the model. We also tested 416 different models consisting of 16 temperatures between $0.5$ keV and $2$ keV in steps of $0.1$ and 26 intrinsic luminosities between $10^{37}$ erg s$^{-1}$ and $10^{42}$ erg s$^{-1}$ in increments of $10^{0.2}$. For both models, we also applied an additional  \texttt{xsphabs} to account for foreground Galactic absorption (Table \ref{Sample}, column 8). 

For each model, we start by generating a spectrum with a long exposure time of unobscured emission and use the flux of the specific model to determine the normalization of the power-law or \texttt{APEC} model such that it would yield the expected flux. We then create a new model by including the obscuration component(s) as appropriate and generate 1000 separate spectra for the exposure taken of the specific galaxy (Table \ref{Sample}, column 11). We used \texttt{sherpa} to generate all of these mock spectra. For each spectrum, we calculate the counts in the 0.5--8 keV, 0.5--2 keV, and 2--8 keV bands, which are saved to a file per model. This process is repeated for each of the 832 models per galaxy and for each galaxy.

We then analyzed these data to generate five maps of the model parameter space. First, we determine the average number of counts and average hardness ratio in the simulated spectra for each model (Figures \ref{workflowPL}a-b and \ref{workflowTH}a-b). The average hardness ratio is determined by calculating the hardness ratio for each spectrum and then averaging the results. These two maps are primarily sanity checks. As expected, low obscuration, high luminosity (lower right in Fig. \ref{workflowPL}a) models of the power-law show high numbers of counts, while high obscuration, low luminosity (upper left) show few counts. For thermal models, in contrast, the number of counts (Fig. \ref{workflowTH}a) depends almost entirely on the luminosity, since the foreground Milky Way obscuration is minor. The hardness ratio plots (Fig. \ref{workflowPL}b and \ref{workflowTH}b) show that, as expected, this quantity broadly does not depend on luminosity, as hardness ratios probe spectral shape, which is primarily determined by the amount of obscuration or the plasma temperature. Hardness ratios tend to be soft for thermal models. 

 Second, we implement our normalization constraint and calculate what fraction of the spectra for each model yielded 0.5--8 keV counts in the range of the observed number for that galaxy (Table \ref{counts}, column 3) plus or minus the Gehrels uncertainty \citep{Gehrels86}, weighted by Poisson probability of observing that many counts if the observed number is the true value\footnote{As an example, SPOG4 has 8 observed counts and we consider spectra with 5 -- 11 counts as within the range consistent with that observation. However, while mock spectra with 8 counts are fully counted, a spectrum with 5 counts would be assigned a 65.6\% weight (the Poisson probability of getting 5 photons per observation from an expected rate of 8).}. This weighting favors models creating large numbers of spectra with the observed number of counts, while taking into account a plausible range consistent with that number due to the uncertainty in how well the observed photon flux represents the true photon flux, given the short finite observation duration as well as the effect of the $\sim$0.3 counts from the background. Examples are shown in Figures \ref{workflowPL}c and \ref{workflowTH}c. The trends seen are however independent of whether we apply this weighting or only count the spectra with the exact number observed.

Third, we undertake a similar calculation for hardness ratio, determining the fraction of spectra for each model that have hardness ratios in the range measured with BEHR for the observation (Table \ref{counts}, column 7; Figures \ref{workflowPL}d, \ref{workflowTH}d). Here we do not apply weighting and instead count all the models within the BEHR range. The reason for doing so is that a large fraction of our galaxies only have 1 count in one of the bands. This is taken into account as part of the BEHR analysis but, as also indicated by the large ranges, means that the median hardness ratios are relatively uncertain. Finally, we combine these two constraints to determine the likelihood that a particular model will yield spectra consistent with the flux and spectral shape of the observations (Figures \ref{workflowPL}e, \ref{workflowTH}e, \ref{ResultsPL}, and \ref{ResultsTH}).

\section{Results}

 \begin{figure*}
\includegraphics[trim=0.1cm 0cm 0.2cm 0.65cm, clip, width=0.5\linewidth]{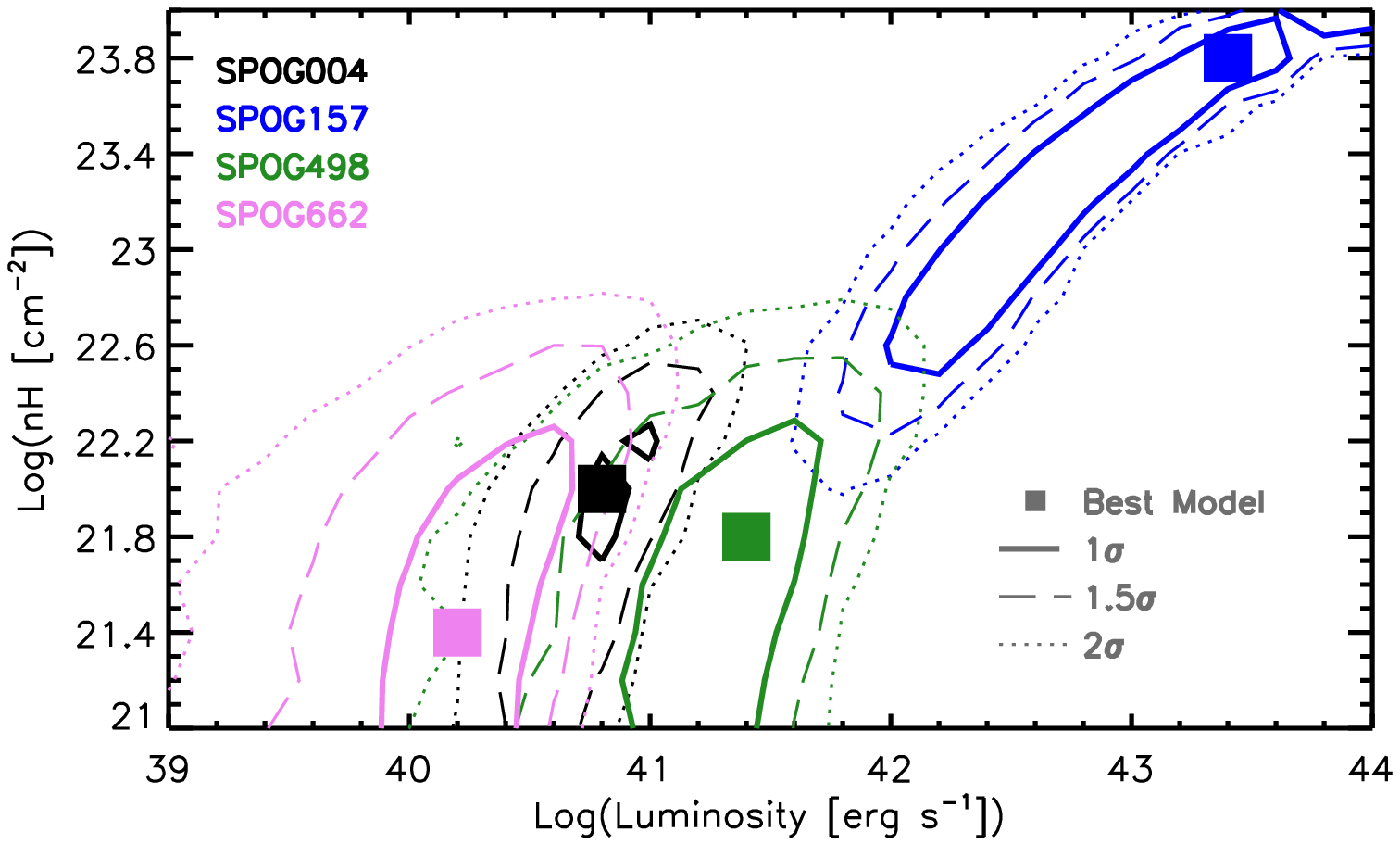}
\includegraphics[trim=0.1cm 0cm 0.2cm 0.65cm, clip, width=0.5\linewidth]{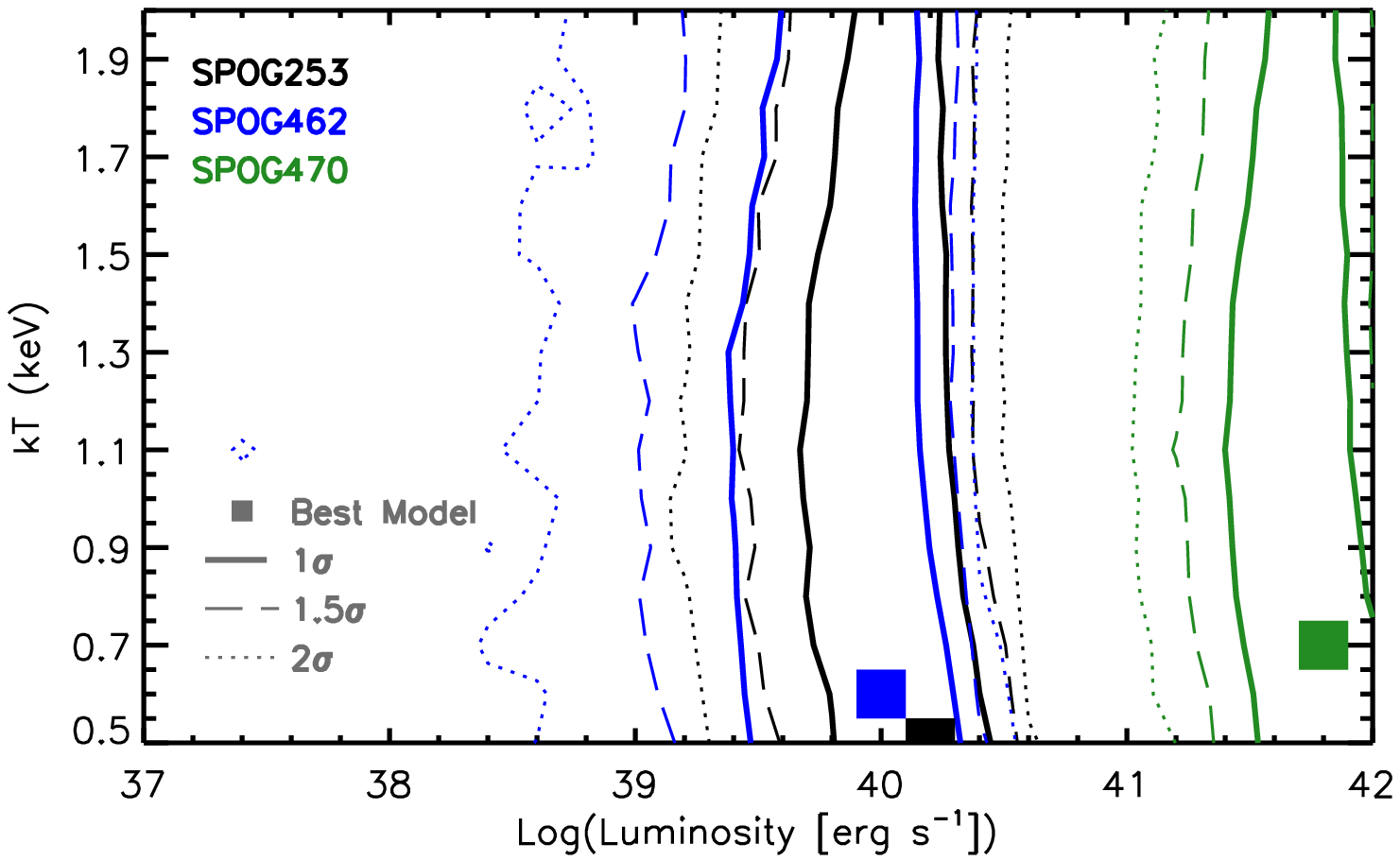}
\caption{ A comparison of the parameter space for the best models for our seven galaxies with significant detections. On the left, we show the four that are significantly better matched with a power-law model. On the right, we show the three models that are better matched with a thermal model. The filled square shows the model with the highest fraction of mock spectra consistent with the observational constraints. The solid, dashed, and dotted lines show the range in which at least 68.3\%, 86.6\%, and 95.4\% (corresponding to 1$\sigma$, 1.5$\sigma$, and 2$\sigma$ of the mean of a normal distribution) of the mock spectra are consistent with the observation. Models outside of each of these contours are rejected at the corresponding level. }
\label{ErrCons}
\end{figure*}

Figures \ref{ResultsPL} and \ref{ResultsTH} show the results of the forward modeling analysis for the power-law and \texttt{APEC} models, respectively. In comparing the results of these models, we see that the galaxies broadly fall into three groups. The first set show much better agreement with one model type or the other: SPOG4 and SPOG157 essentially cannot be reproduced with thermal models, while SPOG253 has much stronger agreement with a thermal model. The second set are not as strongly preferential of one model versus another but have a clear distinction between the highest probability of a thermal vs. a power-law model: SPOG224, SPOG462, and SPOG470 are reproduced more easily with thermal models, while SPOG186, SPOG498, and SPOG662 are more likely to originate from a power-law model. The distinction between these two subgroups is also consistent with their relative hardness ratios (Table \ref{counts}, column 7) as the first subset is distinctly softer than the second. Finally, we have the remaining three galaxies (SPOG305, SPOG689, and SPOG955) where the modeling primarily places some constraints on luminosity and obscuration but does not particularly otherwise narrow the parameter space. These are the three SPOGs with non-significant (even marginally) detections, so it is not altogether surprising that the constraints are weakest in their case.

Table \ref{constraints} lists the parameters of the model with the highest probability and the range of model parameters over which at least 5\% of the simulated spectra are consistent with the observations (corresponding to the dotted line contours in Figure \ref{ErrCons}). For SPOG4, SPOG157, and SPOG253, where one model is clearly better than the other, we only include the parameters of that model. When one model type appears more likely than the other, we bold the model type name. As can be seen in right panels (panel c) of Figure \ref{ResultsPL}, the model parameters of the obscured power-law consistent with the observations are correlated -- i.e., a more obscured source must also be brighter to yield a similar number of counts. 

To illustrate the correlated uncertainties in intrinsic luminosity and obscuration for the power-law models shown in Figures \ref{ResultsPL} and \ref{ResultsTH}, we plot error ellipses in Figure \ref{ErrCons} (left). The best match between model and observed constraint is shown with the filled square. Models outside of the dotted line have less than a 5\% likelihood of resulting in our observation, so they are rejected with 2$\sigma$ confidence. Dashed and solid lines shown similar constraints at 1.5$\sigma$ and 1$\sigma$ respectively. These show that higher intrinsic luminosity models generally require more obscuration to be consistent with our observation, though this is most stringent for SPOG 157. Figure \ref{ErrCons} (right) shows the equivalent figure for the thermal models, demonstrating that while luminosity is generally constrained, temperature is not well constrained for these models.

 \subsection{Constraints on Average Nuclear \\Properties of the Sample}
 
Taken together, these observations and analyses paint a relatively consistent picture regarding the AGN activity in shocked post-starburst galaxies. While two plausible explanations could be given for the low photon flux, low luminosity or high obscuration, the forward modeling analysis indicates that most of these galaxies appear to have at most moderate obscuration and relatively low luminosities. The large majority of the sample have obscurations of $n_H \leq 10^{22}$\,cm$^{-2}$ and intrinsic 2--10 keV luminosities of $L_X \leq 10^{42}$\,erg\,s$^{-1}$. For the significantly detected sources, we can also put a lower limit on the likely luminosity of $L_X \geq 10^{40}$\,erg\,s$^{-1}$, indicating that most of these AGN would be Seyfert-like in their luminosities or low-luminosity AGNs.

\begin{deluxetable*}{lcccccccccc}
\tabletypesize{\scriptsize}
\tablecaption{Parameter Constraints\label{constraints}}
\tablewidth{\textwidth}
\centering
\tablehead{
\colhead{IAU} & \colhead{SPOG} &\colhead{Model} & \multicolumn{2}{c}{Luminosity} &  
\multicolumn{2}{c}{Obscuration} & \multicolumn{2}{c}{Temperature} & \colhead{Notes} \\
\cline{4-5} \cline{6-7} \cline{8-9}
\colhead{Name} & \colhead{Number} & \colhead{Type} & \colhead{Best} & \colhead{Range} &  \colhead{Best} & \colhead{Range} & \colhead{Best} & \colhead{Range}  & \colhead{} \\
\colhead{(1)} & \colhead{(2)} & \colhead{(3)} & \colhead{(4)} & \colhead{(5)} & \colhead{(6)}  & \colhead{(7)}  & \colhead{(8)} & \colhead{(9)}  & \colhead{(10)}  
}
\startdata
\multicolumn{10}{c}{Significantly Detected Galaxies}\\
\hline\\
J001145--005431& ~~~4& {\bf PL} & 40.8  & 40.4 -- 41.2 & 22.0 & $\leq$22.6 &  & & (a) \\ 
J085357+031034 &157  & {\bf PL} & 43.4 & 41.8 -- 44.0 & 23.8 & 22.2 - 24 & & & (a)\\ 
J095750--001239 & 253 & {\bf APEC} & 40.2 & 39.2 -- 40.6 & & & 0.5 & 0.5 -- 2.0 & (b) \\ 
J113655+245325 & 462 & PL & 40.0 & 39.2 -- 40.4 & 21.0 &  $\leq$22.0 &  & & (d) \\
&  & 				    {\bf APEC} & 40.0 & 37.4 -- 40.4 &  &  & 0.6 & 0.5 -- 2.0&  \\ 
J113939+463132 & 470 & PL & 41.8 & 41.4 -- 42.0 & 21.2 &  $\leq$21.8 &  & & (d) \\
&  & 				    {\bf APEC} & 41.8 & 41.2 -- 42.0  &  &  & 0.7 & 0.5 -- 2.0 &   \\ 
J115341+093026 & 498 & {\bf PL} & 41.4 & 40.2 -- 42.0 & 21.8 &  $\leq$22.6 &  & & (c) \\
&  & 				    APEC & 41.2 & 38.6 -- 41.6 &  &  & 2.0 & 0.8 -- 2.0&  \\ 
J131448+210626 & 662 & {\bf PL} & 40.2 & 39.0 -- 41.0 & 21.4 &  $\leq$22.6 &  & & (c) \\
&  & 				    APEC & 40.2 & 39.2 -- 40.4 &  &  & 2.0 & 0.8 -- 2.0 &  \\\\
\hline
\hline
\multicolumn{10}{c}{Marginally Detected or Non-Significantly Detected Galaxies}\\
\hline
J091407+375310 &186  &  {\bf PL} & 41.6 & 39.0 -- 43.2 & 23.4 &  $\leq$24 &  & & (c) \\
&  & 				    APEC & 38 & 37.4 -- 40.6 &  &  & 1.7 & 0.6 -- 2.0 &  \\
J093820+181953 & 224 &  PL & 39.2 & $\leq$41.2 & 21.4 &  $\leq$23.2 &  & & (d) \\
&  & 				    {\bf APEC} & 40.6 & 37.6 -- 41.2 &  &  & 0.6 & 0.5 -- 2.0 &  \\
J102653+434008 & 305 & PL & 39.2 & $\leq$41.2 & 21.2 &  $\leq$23.0 &  & & (e) \\
&  & 				    APEC & $\leq$40 & $\leq$41.2 &  &  &  & 0.5 -- 2.0 &   \\ 
J132648+192246 & 689 & PL & $\leq$41.2 & $\leq$42 & $\leq$22 &  $\leq$22.8 &  & & (e) \\
&  & 				    APEC & $\leq$41.4 & $\leq$41.8 &  &  &  & 0.5 -- 2.0 &   \\ 
J155525+295551 & 955 & PL & $\leq$40.4 & $\leq$41.2 & $\leq$21.8 &  $\leq$22.8 &  & & (e) \\
&  & 				    APEC & $\leq$40.2 & $\leq$41.0 &  &  &  & 0.5 -- 2.0 &   \\ 
\enddata
\tablecomments{(3) Model type described in a row: PL is the power-law model and \texttt{APEC} is the thermal model. (4) Luminosity of the model with the highest probability if models appear well constrain in Figures \ref{ResultsPL} and \ref{ResultsTH}. (5) Range of luminosities over which there are models have at least 5\% of the spectra consistent with the observations. Models outside this range are rejected at the 95\% level. Luminosities are 2--10 keV for power-law models  and 0.5--8 keV for the \texttt{APEC} models. (6-8) Similar constraints on the obscuration column and plasma temperature. (9) Notes on the relative merits of the two models: (a) power-law model is much better, (b) thermal model is much better, (c) the power-law model appears to be better, (d) the \texttt{APEC} model appears to be better, (e) neither model is well constrained.}
\end{deluxetable*}

\subsubsection{The Exceptions: SPOG157 and SPOG186}

Two SPOGs, SPOG157 and to a lesser degree SPOG186,  show a different tendency. SPOG157 requires $n_H \geq 10^{23}$\,cm$^{-2}$, particularly to explain its very hard spectrum. While the model with the highest fraction of mock spectra in agreement with the observed constraints has the parameter pair of $(L_X, n_H) = (3\times10^{43}\,{\rm erg\,s^{-1}}, 6\times10^{23}\,{\rm cm^{-2}})$, the six models in Figure \ref{ResultsPL}c for SPOG157 with the darkest blue color have a similar fraction of mock spectra in agreement with the observations. They extend from  $(L_X, n_H) = (3\times10^{42}\,{\rm erg\,s^{-1}}, 1\times10^{23}\,{\rm cm^{-2}})$ to  $(L_X, n_H) = (6\times10^{43}\,{\rm erg\,s^{-1}}, 1\times10^{24}\,{\rm cm^{-2}})$, indicating that SPOG157 is more luminous and more obscured than most of the sample. 

SPOG186 is only marginally detected, so its constraints are weaker. It too has a best fit model with a higher obscuration, albeit not as high of a luminosity, of $(L_X, n_H) = (4\times10^{41}\,{\rm erg\,s^{-1}}, 3\times10^{23}\,{\rm cm^{-2}})$. Its set of six models with similarly high fractions of mock spectra in agreement with observed constraints has a parameter pair range extending from  $(L_X, n_H) = (3\times10^{41}\,{\rm erg\,s^{-1}}, 2\times10^{23}\,{\rm cm^{-2}})$ to  $(L_X, n_H) = (4\times10^{42}\,{\rm erg\,s^{-1}}, 1\times10^{24}\,{\rm cm^{-2}})$. 

Given that the range of similarly-well fit models for both of these extends up to the $n_H = 1 \times10^{24}\,{\rm cm^{-2}}$ limit of the model parameters, it is possible that one or both could have Compton-thick obscuration. However, the absorption model used, \texttt{XSPEC}'s photoelectric absorption (\texttt{phabs}; \texttt{xsphabs} in \texttt{sherpa}) ceases to apply when the obscuration becomes optically thick at $n_H \gtrsim 1\times10^{24}\,{\rm cm^{-2}}$ \citep[e.g.,][]{Maiolino07}. Testing parameters in this range would require a more complex model  (e.g., \texttt{BORUS}; \citealt{Balokovic18}), which have more free parameters, thereby making them difficult to constrain with the limited data available for these galaxies.

\section{Discussion}

 \subsection{AGNs in Transitioning Galaxies}
 
The overarching question this study sought to help address was the role of AGNs in the transition of galaxies from actively star-forming to quiescence. The X-ray observations we have taken suggest that a large fraction ($\approx 58 - 75\%$) appear to have at least a moderate degree of nuclear activity. What is clear though is that we are not detecting bright obscured or unobscured quasars, as in the major-merger evolutionary scenario \citep[e.g.,][]{sanders88, hopkins08, alexander12}. As such, it is clear that these AGN are not capable of radiatively driving out the star-forming fuel and driving the transition to quiescence. Rather, it seems that the mechanisms that started them on this transition are likely to have directed gas to the centers of these galaxies and thereby fed the accretion flow of these supermassive black holes. 

\citet{Sazonova21} recently undertook a morphological analysis of 26 SPOGs observed with the {\em Hubble Space Telescope}, finding that 27\% show clear evidence of recent merger activity and another 30\% show disturbances. Due to the fading of merger signatures over time, it is difficult to conclude whether all disturbances are the results of merger activity or whether alternative internal processes such as AGN-driven outflows may cause at least some of them. 

Six galaxies were common between that subsample of SPOGs and those studied here. Interestingly, both SPOG157 and SPOG186 appear to be edge-on dusty disk galaxies, suggesting that at least some part of their obscuration may be on the galactic scale rather than purely at the nuclear scale. The other four galaxies (SPOG4, SPOG224, SPOG253, and SPOG305) show disturbed but bulge-dominated morphologies without tidal features that would indicate a clear tidal origin. On the X-ray front, these four galaxies span the range of hardness ratios (excluding SPOG157 and SPOG186) and are representative of the luminosity range and model types of the other 6 galaxies. As such, similar processes may be generating the X-ray emission in all these galaxies. 

The presence of morphological disturbances indicates either the likely presence of mechanisms that could bring gas to the supermassive black holes turning them into modest AGNs or the effects of current or recent AGN mechanical feedback driving an outflow or turbulence into the host galaxy generating signatures of disturbances. In the former case, the active AGN would manifest as the power-law emission in X-rays (e.g., SPOG4 or SPOG498). In the latter case, AGN-driven outflows are often multiphase and can include hot X-ray emitting plasma \citep[e.g., NGC\,1266;][]{alatalo15}, but disentangling the relative contributions of thermal and power-law emission would be difficult without making further assumptions to reduce some of the free parameters of the combined model, given the limited constraints of the observations.

SPOG253, SPOG462, SPOG470, and SPOG224 all seem to have similar observed properties, namely X-ray emission dominated by thermal emission at temperatures around $0.5-0.7$\,keV with luminosities generally of a similar order of magnitude ($L_X\sim10^{40}$\,erg\,s$^{-1}$), although SPOG470 looks to be a bit brighter. The relatively common presence of thermal emission in this sample suggests that  either  multiphase outflows or shocked X-ray emitting plasma, potentially due to interactions between AGN jets and ISM (e.g., \citealt{ogle14, lanz15, appleton18}),  may also be frequent in SPOGs.  

Since thermal emission can also be associated with X-ray binaries and star formation, we tested the likelihood of this possibility by calculating the expected X-ray luminosity under the assumption that all of the H$\alpha$ emission in the SDSS fiber\footnote{DR7 spectra were taken through 3$''$ apertures, slightly larger than our X-ray extraction aperture.} was due to star formation. We calculated the associated star formation rate with \citet{Kennicutt94} relation and the associated X-ray binary luminosity using Equation 22 from \citet{Mineo12}. For all four of the galaxies where thermal models appear preferred, the X-ray binaries luminosity is at least an order of magnitude less than the most likely X-ray luminosity from the forward-modeling, indicating that the X-ray emission is unlikely to be solely due to X-ray binaries. 

However, additional multiwavelength observations and analyses would be needed to confirm this interpretation and rule out nuclear starbursts as an alternative source. One possible avenue would be mid-infrared spectra to search for warm molecular hydrogen emission lines indicative of shocked ISM. If these galaxies also have deeply buried AGN responsible for outflows like NGC\,1266 or other AGN-driven shocks, then at least 75\% of the sample would contain AGN of modest to moderate luminosity and therefore accretion rate.

While these AGN are not radiatively powerful, this frequency of X-ray activity is high compared to extragalactic populations in X-ray surveys. For field galaxies, as these SPOGs appear to generally be, \citet{haggard10} found that approximately 1\% at redshifts $0.05 < z < 0.31$ had AGN with $L_X\geq 10^{42}$\,erg\,s$^{-1}$. Only SPOG157 meets this luminosity threshold, so the small number statistics make it difficult to tell whether bright AGN activity is over-represented in SPOGs at 8\% (1/12). However, low-luminosity AGN activity does seem to be higher than found in other extragalactic populations. \citet{young12} investigated the presence of low-luminosity AGN via variability in the 92 galaxies ($0.08 < z < 1.02$)  in the 4 Ms Chandra Deep Field South (CDF-S) that do not contain AGN with $L_X>3\times10^{42}$\,erg\,s$^{-1}$. They found that at least 20 ($\geq22$\%) showed evidence of accretion onto a supermassive black hole, indicating that growth in this quiescent or low-luminosity phases could contribute significantly to black hole mass. 

By comparison, the $50-75$\% of our SPOGs showing at least low-luminosity AGN emission is suggestively higher\footnote{Using Poisson statistics, the cumulative probability that at least 6/12 in our sample show low-luminosity AGN signatures if they belong to the same population as the CDF-S set is 5.2\%.}. This is consistent with the previous results such as those of \citet{Depropris14}, in finding that X-ray emission is common but indicative of low-luminosity AGN activity rather than radiatively powerful AGN. Further observations will however be required to better constrain the black hole and galactic bulge masses to see if growth was potentially needed in this transitional stage to bring the galaxies in agreement with the M$-\sigma$ relation \citep[e.g.,][]{mcconnell13}.

\subsection{Implication for Timing of AGN Activity}

The canonical picture of galaxy evolution via major mergers (e.g., \citealt{sanders88, hopkins06}) places AGN feedback prior to a post-starburst phase, requiring strong radiation from the AGN to drive out star forming fuel. While this pathway may work well for some galaxies, others are likely to require different relative timings on these processes. For example, a significant fraction of two post-starburst samples  showed significant remaining reservoirs of molecular gas \citep{rowlands15, alatalo16b_spog2, french18} and \citet{schawinski09} found similar reservoirs powering residual star formation in a sample of early-type galaxies. The galaxies in these samples require a mechanism to either complete the removal of this gas or otherwise suppress its ability to form stars. \citet{schawinski09} specifically concluded that low-luminosity AGN episodes were the most likely method by which this would be accomplished in their sources. 

A delay between the starburst phase and the onset of AGN activity would be in line with a number of other studies. \citet{wild07, wild10} and \citet{davies07} found such delays in samples of low-redshift AGNs, many of which also had post-starburst signatures. \citet{yesuf14} and \citet{pawlik18} probed this question starting with post-starburst galaxies and found AGN signatures in a significant fraction. However, both studies also concluded that while AGN may prevent re-emergence of star formation activity or help finish quenching, the later onset indicated it was unlikely that the AGN had precipitated the primary decrease in star formation activity. 

This pathway with delayed AGN activity would explain our particular galaxies better than the canonical picture. While an earlier phase of luminous AGN activity may have taken place at the peak of any major merger that these galaxies may experienced, the current AGN activity is much more consistent with lower luminosities that can act to suppress residual star formation activity.

 \subsection{Effects of Sample Selection \label{bias}}
 
 In selecting the subset of SPOGs to observe with {\em Chandra}, we did not select our pilot sample to be representative of the full SPOG population. Instead, we focused on a set with the most complete datasets. As such, our set of 12 had been detected in CO(1--0) \citep{alatalo16b_spog2} with either IRAM or CARMA. SPOGs selected for these submm observations had in turn been required to be detected with WISE in the 22\um band (491/1067 SPOGs). While not all of the CO-observed SPOGs also had FIRST counterparts, we selected our Chandra targets from the subset that did.

This sample selection has several implications for the broader interpretation of what this sample can tell us about the population. First, the requirement of a CO detection means that these galaxies will still contain potential fuel for AGN activity. Post-starburst galaxies have been shown to have molecular gas fractions more similar to star-forming galaxies than to the quiescent elliptical and lenticular galaxies they are thought to be evolving into \citep{french15, rowlands15, alatalo16b_spog2}. However, a mechanism such as a merger is likely necessary to disrupt the star-forming disk and help drive the gas towards the nucleus in order to feed the AGN \citep[e.g.,][]{hopkins12}. 

Second, the requirement of both a significant 22\um detection and a FIRST counterpart selected a pilot sample with pre-existing indications of AGN activity. As such, these SPOGs may be some of those most likely to host an active AGN and the AGN fraction in this sample may therefore be higher than in the overall SPOG population. However, infrared, radio, and X-ray AGN selection criteria do not select the same AGN \citep{hickox09}, suggesting that radio-bright AGN may not necessarily be X-ray or infrared bright. Indeed, \citet{hickox09} found that only $\sim10$\% of radio-selected AGN were also selected in the infrared and of those, only about half were also selected in X-rays. As such, it is possible that brighter X-ray AGN exist within the set of SPOGs without FIRST counterparts. Extending the conclusions of this study to the larger post-starburst population including E+A and K+A galaxies is also complicated by the relatively younger age of SPOGs (see Fig. 9 of \citealt{alatalo16a_spog1}) and the unclear timing of AGN triggering during the transition to quiescence. 

Third, the FIRST counterpart requirements may also indicate that, as radio AGN,  the feedback mode of the AGN in these galaxies will tend to the mechanical rather than radiative \citep[e.g.,][]{alexander12}. As such, the AGN could still be important in helping the transition along. NGC\,1266 shows that SPOGs can contain significant AGN-driven outflows \citep{alatalo15}. However, it also shows that, while such outflows can be effective at injecting turbulence into the ISM and pushing the gas outwards from the nuclear region, these outflows are unlikely to fully eject the gas reservoirs out of the gravitational potential of these galaxies. Therefore, the AGN in these systems may be better said to be helping maintain a lower level of star-forming activity by reducing the amount of cold dense gas rather than driving a permanent end to star formation as would be the case with a removal of the remainder of the star-forming fuel. The significant fraction of the sample with X-ray emission consistent with thermal emission suggests that this may be a frequent process in these galaxies. 
 
\subsection{Future Work}

Addressing the effects of the sample selection in order to generalize these conclusions to the broader SPOG and post-starburst population will require larger samples of X-ray observations. As we have shown in this paper, the forward modeling methodology allows us to place constraints even with minimal photons. The next step is clearly to apply this approach to all observations of post-starburst galaxies observed, primarily serendipitously, with {\em Chandra}, {\em XMM-Newton}, {\em Swift}-XRT, {\em NuSTAR}, and eROSITA. The large sky coverage of eROSITA is particularly promising as a way of reducing the effects of selection biases, and extending to higher energies could give improved constraints on obscured systems like SPOG157 and/or enable us to determine whether systems that are thermally dominated at softer energies (e.g., SPOG253, NGC\,1266) also have deeply buried AGN. 

A second area of improvement on which we are actively working is to determine how firm the forward modeling constraints are. Our conclusions in this paper do not depend on the specific luminosity or obscuration of the model that matches best. But it would be useful to know how strong those constraints are and whether this approach tends to systematically over- or under-estimate certain parameters. We are taking two approaches to testing this methodology. First, we will use new deeper {\em XMM-Newton} observations on SPOG157 (70\,ks) and SPOG253 (63\,ks). These should provide sufficient counts for spectral fitting, providing an independent constraint on the parameters of these two SPOGs to compare with the forward modeling analysis. Further, as the two galaxies from this sample span the range of hardness ratios and model types, those analyses may also provide further guidance on the forward-modeling methodology. Second, we are applying the forward modeling analysis to a set of {\em Swift}-BAT sources, whose parameters are well constrained by a combination of {\em Swift}-XRT and {\em NuSTAR} observations to see how well the forward modeling analysis recovers their luminosity and obscuration. That analysis will show the reliability of the forward modeling constraints.

\section{Summary}

We performed a forward modeling analysis for 12 post-starburst galaxies from the SPOGs sample to constrain the parameters of their nuclear X-ray emission. At least 7 of the 12 had significant X-ray detections in $\sim$10\,ks {\em Chandra} observations. However, none of these sources had more than 10 photons. We used the photon flux and the range of hardness ratios consistent with the observations to constrain the intrinsic 2--10\,keV luminosity and obscuration column of a simple AGN model or the 0.5--8\,keV luminosity and temperature of an APEC thermal model. For each galaxy, we tested 832 different models for which we generated 1000 mock spectra each to compare with the observed constraints. 

Between 58\% and 75\% of the sample show evidence of AGN activity, depending on whether we include galaxies with weaker constraints, which could also be the result of thermal emission. The large majority of these are low luminosity AGNs and even the brightest, SPOG157, only has a luminosity of $\sim3\times10^{43}$\,erg\,s$^{-1}$. Taken together, these analyses suggest that a large fraction of supermassive black holes in SPOGs are active, especially compared to the broader population of field galaxies. However, they are clearly not sufficiently luminous to exert significant radiative feedback. 

The selection criteria of this pilot sample may have selected both for and against X-ray bright AGN (see Section \ref{bias}), so how well this conclusion can be extended to the larger SPOG sample  and the broader post-starburst population remains unclear. The general SPOGs selection tends to select galaxies with a younger stellar population, caught earlier in the transitional process, and its basis on line ratios rather than line strengths does not exclude as many galaxies that have line properties consistent with AGN or LINER emission. The resulting higher level of activity in these galaxies may reveal its importance to the initial phases of post-starburst transition, though it may not  remain as prevalent throughout the post-starburst phase. FIRST detections of each of this sample's galaxies may also indicate that the mechanical mode of feedback is more common in this phase.  Future analyses of existing and forthcoming X-ray datasets, as well as radio and infrared signatures of AGN, may provide the greater clarity the larger question of the timing and activity of AGN in the full post-starburst phase. 

\begin{acknowledgements}
L.L. thanks Alexander Nazarov for his contributions in helping to lay the groundwork for this analysis and Michael Ochs for useful discussions on Bayesian statistics. The scientific results reported in this article are based on observations made by the Chandra X-Ray Observatory. Support for this work was provided by the National Aeronautics and Space Administration through Chandra Award Number GO7-18093A issued by the Chandra X-Ray Observatory Center, which is operated by the Smithsonian Astrophysical Observatory for and on behalf of the National Aeronautics Space Administration (NASA) under contract NAS8-03060. L. L. and S.S. also acknowledge support from NASA through grant number 80NSSC20K0050. Basic research in radio astronomy at the U.S. Naval Research Laboratory is supported by 6.1 Base Funding. AMM acknowledges support from the National Science Foundation under Grant No. 2009416.
\end{acknowledgements}

\facility{CXO}
\software{Sherpa \citep{sherpa,sherpaPython}}
\bibliography{bib_used.bib}

\end{document}